\documentclass[fleqn,usenatbib]{mnras}
\usepackage{newtxtext,newtxmath}
\usepackage[T1]{fontenc}
\usepackage{romannum}
\usepackage{graphicx} % Including figure files
\usepackage[tight,footnotesize]{subfigure}
\usepackage{amsmath}	% Advanced maths commands
\usepackage{multirow}
\usepackage{comment}
\usepackage{textcomp}
\hypersetup{
    colorlinks=true,
    linkcolor=blue,
    filecolor=blue,      
    urlcolor=magenta,
    citecolor=blue,
}
\newcommand{\orb}{${\Omega}$\,} % Omega
\newcommand{\twoorb}{${2\Omega}$\,} % 2Omega
\newcommand{\threeorb}{${3\Omega}$\,} % 2Omega
\newcommand{\po}{P$_{\Omega}$\,} % P_Omega
\newcommand{\ptwoo}{P$_{2\Omega}$\,} % P_2Omega
\newcommand{\pthreeo}{P$_{3\Omega}$\,} % P_3Omega
\newcommand{\rn}[1]{%
  \textup{\uppercase\expandafter{\romannumeral#1}}%
}
\title[Two Polars: J1743 and YY Sex]{Confirmation of two magnetic cataclysmic variables as polars: 1RXS J174320.1-042953 and YY Sex}

\author[Rawat et al.]
{
Nikita Rawat $^{1,2}$\thanks{E-mail: nikita@aries.res.in, rawatnikita221@gmail.com},
J. C. Pandey $^{1}$, 
Arti Joshi $^{3}$,
Stephen B. Potter $^{4,5}$,
Alisher S. Hojaev $^{6}$,
Micha\"{e}l De Becker $^{7}$, 
\newauthor
\hspace{0.1cm}Srinivas M Rao $^{1}$ and 
Umesh Yadava $^{2}$
\\
$^{1}$Astronomy Division, Aryabhatta Research Institute of observational sciencES (ARIES), Nainital 263001, India\\
$^{2}$Deen Dayal Upadhyaya Gorakhpur University, Gorakhpur 273009, India\\
$^{3}$ Indian Institute of Astrophysics (IIA), Koramangala, Bangalore 560034, India\\
$^{4}$ South African Astronomical Observatory, PO Box 9, Observatory, Cape Town 7935, South Africa\\
$^5$ Department of Physics, University of Johannesburg, PO Box 524, 2006 Auckland Park, Johannesburg, South Africa\\
$^{6}$ Ulugh Beg Astronomical Institute, Uzbekistan Academy of Sciences, Tashkent, Uzbekistan\\
$^{7}$ Space Sciences, Technologies and Astrophysics Research (STAR) Institute, University of Li\`{e}ge, Quartier Agora, 19c, All\'{e}e du 6 A\^{o}ut, B5c, B-4000 Sart Tilman, \\
Belgium\\
}

\pubyear{2022}
\date{Accepted XXX. Received YYY; in original form ZZZ}

\begin{document}
\label{firstpage}
\pagerange{\pageref{firstpage}--\pageref{lastpage}}
\maketitle

\begin{abstract}
We present our analysis of new and archived observations of two candidate magnetic cataclysmic variables, namely 1RXS J174320.1-042953 and YY Sex. 1RXS J174320.1-042953 was observed in two distinctive high and low states where a phase shift was seen, which could be due to the changes in the shape, size, and (or) location of the accretion region. We find that its orbital X-ray modulations only persist in the soft (0.3-2.0 keV) energy band, which could be attributed to the photoelectric absorption in the accretion flow.  The X-ray spectra exhibit a multi-temperature post-shock region where the hard X-rays are absorbed through a thick absorber with an equivalent hydrogen column of $\sim$7.5 $\times$ 10$^{23}$ cm$^{-2}$, which partially covers $\sim$56 per cent of the emission. No soft X-ray excess was found to be present; however, a soft X-ray emission with a blackbody temperature of $\sim$97 eV describes the spectra. Extensive \textit{TESS} observations of YY Sex allow us to refine its orbital period to 1.5746 $\pm$ 0.0011 h. We did not find any signature of previously reported spin or beat periods in this system. Furthermore, our new polarimetric observations show clear circular polarization modulated on the orbital period only. Finally, both systems show strong Balmer and He \rn{2} 4686 \AA  ~emission lines in the optical spectra, further indicative of their magnetic nature.

\end{abstract}

\section{Introduction} \label{sec1}

Magnetic cataclysmic variables (MCVs) are interacting semi-detached binaries consisting of a magnetic white dwarf (WD) as the primary and a Roche lobe filling star as the secondary, which loses material to the primary through the inner Lagrangian point. The magnetic field strength of the WD plays a crucial role  in deciding the two distinct subclasses of MCVs: intermediate polars (IPs) and polars. IPs are asynchronous systems in which  the magnetic field strength of the WD is relatively weaker (B $\sim$1-10 MG) and an accretion disc can form, which truncates at the magnetospheric radius \citep[see][for a full review of IPs]{1994PASP..106..209P}. 
%Two prominent periods (orbital period of binary system and spin period of WD) differing by  $\sim$10 per cent in values is a characteristic feature of this subclass \citep[see][for a full review of IPs]{1994PASP..106..209P}.
On the other hand, polars have a comparatively higher magnetic field than IPs, which prevents the formation of an accretion disc. It also keeps both stars in a synchronous rotation.
%i.e. the orbital period of the binary system (\po) = spin period of WD (\ps). 
Polars have typical periods below the period gap of 2-3 h \citep{2010MNRAS.401.2207S}. The mass transfer in polars occurs along the magnetic field lines to the WD magnetic pole(s) \citep{1990SSRv...54..195C,1995cvs..book.....W}. As the accreting matter falls supersonically towards the WD surface, it produces a shock-front, which heats the infalling plasma to several tens of keV. The flow cools down via thermal plasma radiation at X-rays \citep{1995cvs..book.....W} and cyclotron radiation at infrared, optical, and ultraviolet wavelengths \citep{1973PThPh..49.1184A}. The emitted radiation is mostly reflected for E $\geq$30 keV from the WD surface. However, the lower energies are absorbed, thermalized, and re-emitted as a blackbody in the UV or soft X-ray region with a blackbody temperature of $\sim$40 eV \citep{1995cvs..book.....W}. Therefore, the X-ray spectra of polars are well described with multi-temperature plasma components. While the optical spectrum of typical polar exhibits strong Balmer emission lines, He \rn{1}, He \rn{2}, and C \rn{3}/N \rn{3} blend at 4650 \AA \hspace{0.05cm} originating in the accretion column. Further, the strengths of He \rn{2} 4686 \AA ~and H$\beta$ are comparable \citep{1995cvs..book.....W}. 
%Due to the absence of accretion discs, the spectra of polars usually present narrow single-peaked emission lines, which are generally asymmetric since they are formed in different regions in the accretion flow. 
Because of the lack of accretion discs, the spectra of polars typically contain narrow single-peaked emission lines that are generally asymmetric because they form in different regions in the accretion flow. In this paper, we present detailed analyses of two candidate MCVs, namely 1RXS J174320.1-042953 \citep[=V3704 Oph;][]{2019IBVS.6261....1K} and YY Sex.

%Table-1
\begin{table*}
\begin{center}
\caption{A log of optical photometric, spectroscopic, photo-polarimetric,  and X-ray observations of J1743 and YY Sex.}
\label{tab:obslog}
\end{center}
\renewcommand{\arraystretch}{1.2}
\begin{tabular}{lcccccc}
\hline
Object & Date of & Telescope & Instrument & Filter/Band & Exposure & Integration \\
& Observations & & & & Time (s) & Time (h) \\
\hline
J1743 & 2019 Sep 21 & \textit{XMM-Newton} & EPIC and OM & 0.3-10.0 keV & 21000 & 5.83 \\
& 2021 Mar 16 & 1.3m-DFOT & 2k $\times$ 2k CCD & R & 150 & 2.76 \\
& 2021 Mar 17 & 1.3m-DFOT & 2k $\times$ 2k CCD & R & 150 & 2.92 \\
& 2021 May 28 & 2m-HCT & HFOSC/Gr7 & 380-684 nm & 2700 & 0.75\\
& 2021 May 29 & 0.6m Zeiss-600 Northern & 1k $\times$ 1k CCD & R & 240 & 3.34\\
& 2021 May 30 & 0.6m Zeiss-600 Northern & 1k $\times$ 1k CCD & R & 240 & 2.95\\
& 2021 May 31 & 0.6m Zeiss-600 Northern & 1k $\times$ 1k CCD & R & 240 & 1.59\\
& 2021 June 06 & 1.3m-DFOT & 2k $\times$ 2k CCD & R & 60 & 4.81\\
& 2021 June 08 & 1.5m AZT-22 & 4k $\times$ 4k CCD & R & 180 & 2.18\\
& 2021 June 10 & 1.5m AZT-22 & 4k $\times$ 4k CCD & R & 180 & 2.23\\
& 2021 June 11 & 1.5m AZT-22 & 4k $\times$ 4k CCD & R & 180 & 1.68\\
& 2021 June 14 & 1.5m AZT-22 & 4k $\times$ 4k CCD & R & 240 & 2.24\\
& 2022 Apr 09 & 1.3m-DFOT & 2k $\times$ 2k CCD & R & 120 & 2.59\\

\hline
YY Sex   & 2004 May 22 & \textit{XMM-Newton} & EPIC and OM & 0.3-10.0 keV & 23954 & 6.65 \\
 & 2004 June 21 & \textit{XMM-Newton} & EPIC and OM & 0.3-10.0 keV & 8911 & 2.48 \\
 & 2010 Mar 12 & SAAO 1.9m & HIPPO & OG57 & 0.01 & 5.76 \\
& 2021 Feb 09 & \textit{TESS} & Photometer & 600-1000 nm & 120 & 579.84 \\

 & 2021 Dec 14 & 1.3m-DFOT & 2k $\times$ 2k CCD & R & 240 & 2.25 \\
& 2022 Jan 24 & 2m-HCT & HFOSC/Gr7 & 380-684 nm & 3600 & 1.00\\
 & 2022 Feb 07 & 1.3m-DFOT & 2k $\times$ 2k CCD & R & 300 & 2.41 \\
 & 2022 Feb 11 & 1.3m-DFOT & 2k $\times$ 2k CCD & R & 300 & 5.03 \\
 & 2022 Mar 25 & 1.3m-DFOT & 2k $\times$ 2k CCD & R & 300 & 4.15 \\
 & 2022 Mar 26 & 1.3m-DFOT & 2k $\times$ 2k CCD & R & 300 & 3.20 \\
 & 2022 Mar 31 & 1.3m-DFOT & 2k $\times$ 2k CCD & R & 300 & 3.38 \\
 & 2022 Apr 08 & 1.3m-DFOT & 2k $\times$ 2k CCD & R & 300 & 2.19 \\
 & 2022 Apr 09 & 1.3m-DFOT & 2k $\times$ 2k CCD & R & 300 & 2.94 \\

\hline
\end{tabular}
\end{table*}

\par 1RXS J174320.1-042953 (hereafter J1743) was identified as a CV by \citet{2011AstL...37...91D}. \citet{2012PZ.....32....3D} performed white-light photometry and suggested that J1743 is polar with an orbital period (\po) of 2.078(7) h. However, the small fluctuations in the light curve led them to suspect that J1743 is an IP with a quickly spinning WD. The unique features of the light curve they obtained were the large amplitude (0.8 mag) oscillations and a dip by about 0.4 mag shortly before the maximum. \citet{2017AJ....153..144O} suggested J1743 to be polar based on the optical spectrum, where they found intense He \rn{2} 4686 \AA, C \rn{3}/N \rn{3}, ~and H$\beta$ more intense than H$\alpha$. It is located at a distance of 214 $\pm$ 2 pc \citep{2021AJ....161..147B}.

\par YY Sex (discovery name `RX J1039.7-0507') was identified as a CV by \citet{1998ApJS..117..319A} based on the emission line spectrum. \citet{2003MNRAS.339..731W} proposed YY Sex to be an IP, having claimed the detection of an orbital period of 1.574 h, a spin period of 1444 s, and a beat period of 1932.5 s in their optical power spectrum. In addition, they interpreted the nearly sinusoidal modulation in their optical light curves with large amplitude variations (1.1 mag) as the reflection effect caused by a very hot WD. Further, \cite{2017ASPC..510..435G} performed optical spectroscopy and suggested that YY Sex is polar with the presence of strong hydrogen Balmer H $\beta$, H$\gamma$, H$\delta$, He \rn{1}, and He \rn{2} 4686 \AA ~lines. They also showed, using doppler mapping, that there is no sign of disc  accretion. The Gaia distance of this source is 382$^{+31}_{-22}$ pc \citep{2021AJ....161..147B}.

\par Prior studies have been ambiguous about the nature of these two sources. Therefore, with a motivation to ascertain their true nature, we selected these two sources for their detailed analyses. 
This paper is organized as follows. The following section summarizes observations and data reduction. Section  \ref{sec3} includes the analyses and results of the optical and X-ray data. Finally, in Sections \ref{sec4} and \ref{sec5}, we present the discussion and conclusions, respectively.
%We summarize observations and data reduction  in the next section. Section \ref{sec3} contains analyses and the results of the optical and X-ray data. Finally, we present the discussion and conclusions in Sections \ref{sec4} and \ref{sec5}, respectively.

\section{Observations and Data Reduction} \label{sec2}

\subsection{Optical photometric observations} \label{sec2.1}
R-band photometric observations of these sources were acquired in 2021 and 2022 using 1.3-m Devasthal Fast Optical Telescope (DFOT) located at Devasthal, Nainital, India \citep{2011CSci..101.1020S}, and  1.5-m (AZT-22) and 0.6 m Zeiss-600 Northern telescopes at Maidanak observatory Uzbekistan \citep{2010ARep...54.1019A}.  A detailed log for photometric observations is given in Table \ref{tab:obslog}. The 1.3-m DFOT has Ritchey-Chretien (RC) design with an f/4 beam at the Cassegrain focus. It is equipped with a 2k$\times$2k Andor CCD, with each pixel having a size of 13.5 $\mu \rm m^{2}$, which covers a total field of view of    $\sim$18$\arcmin \times$ 18$\arcmin$. At 1 MHz read-out speed, the read-out noise and gain are 7.5 e$^{-}$/pixel and 2 e$^{-}$/ADU, respectively. The 1.5-m AZT-22 also has an RC design. We used  4k$\times$4k CCD (SNUCAM) with read-out noise and gain of 4.7 e$^{-}$/pixel and 1.45 e$^{-}$/ADU at 200 kHz read-out speed, respectively. J1743 was also observed by the 0.6-m Zeiss-600 telescope with an f/12.5 beam at the Cassegrain focus. It was equipped with a 1k$\times$1k CCD with a read-out noise of 13 e$^{-}$/pixel and gain of 5 e$^{-}$/ADU. 

\par During the observing runs, several bias and twilight sky flat frames were taken. \textsc{iraf}\footnote{\textsc{iraf} is distributed by the national optical astronomy observatories, USA.} routines were used to perform the pre-processing steps like bias subtraction, flat-fielding, and cosmic-ray removal of raw data. Using a comparison star in the same field, differential photometry (variable minus comparison star) was performed. Table \ref{tab:detailslog} provides information on comparison stars (C1 and C2) for each source. The R-band magnitudes of both sources were calculated with respect to comparison star C1. The nightly variations ($\sigma$) of C1-C2 were found to be between 0.005-0.02 (for J1743) and 0.009-0.019 (for YY Sex). The USNO-B1.0 catalogue served as the source for the standard magnitudes of C1 and C2  (see Table \ref{tab:detailslog}).

%Table-2
\begin{table}
\begin{center}
\caption{Comparison stars used for the differential photometry for J1743 and YY Sex.}
\label{tab:detailslog}
\end{center}
\renewcommand{\arraystretch}{1.0}
\begin{tabular}{lccccc}
\hline
Object & Reference & B1 & R1 & B2 & R2  \\
& USNO-B1.0 & (mag) & (mag) & (mag) & (mag)  \\
\hline
J1743 & 0855-0326594 & 16.12 & 15.15 & 18.95 & 18.13 \\
C1 & 0854-0330352 & 17.59 & 15.63 & 17.10 & 15.75 \\
C2 & 0855-0326373 & 18.13 & 15.75 & 17.15 & 15.82 \\
\hline
YY Sex & 0848-0214615 & 19.09 & 18.83 & 18.50 & 19.32\\
C1 & 0848-0214689 & 18.40 & 16.22 & 18.65 & 16.25\\
C2 & 0849-0213686 & 18.10 & 16.34 & 18.26 & 16.61\\
\hline
\end{tabular}

\bigskip
\emph{\textbf{Note.}} C1 and C2 stand for comparisons 1 and 2 for each target star.
\end{table}

% Figure-1
\begin{figure*}
\centering
\subfigure[]{\includegraphics[width=0.85\textwidth]{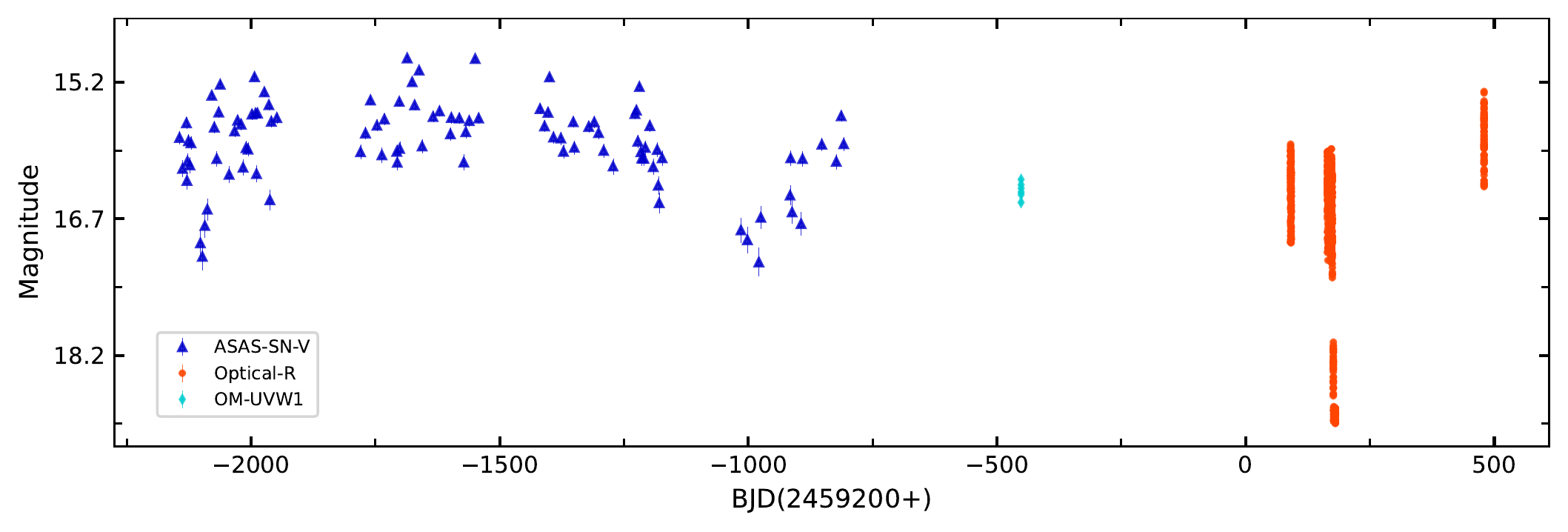} \label{fig:J1743_lc_all}}
\subfigure[]{\includegraphics[width=0.85\textwidth,height=8.5cm]{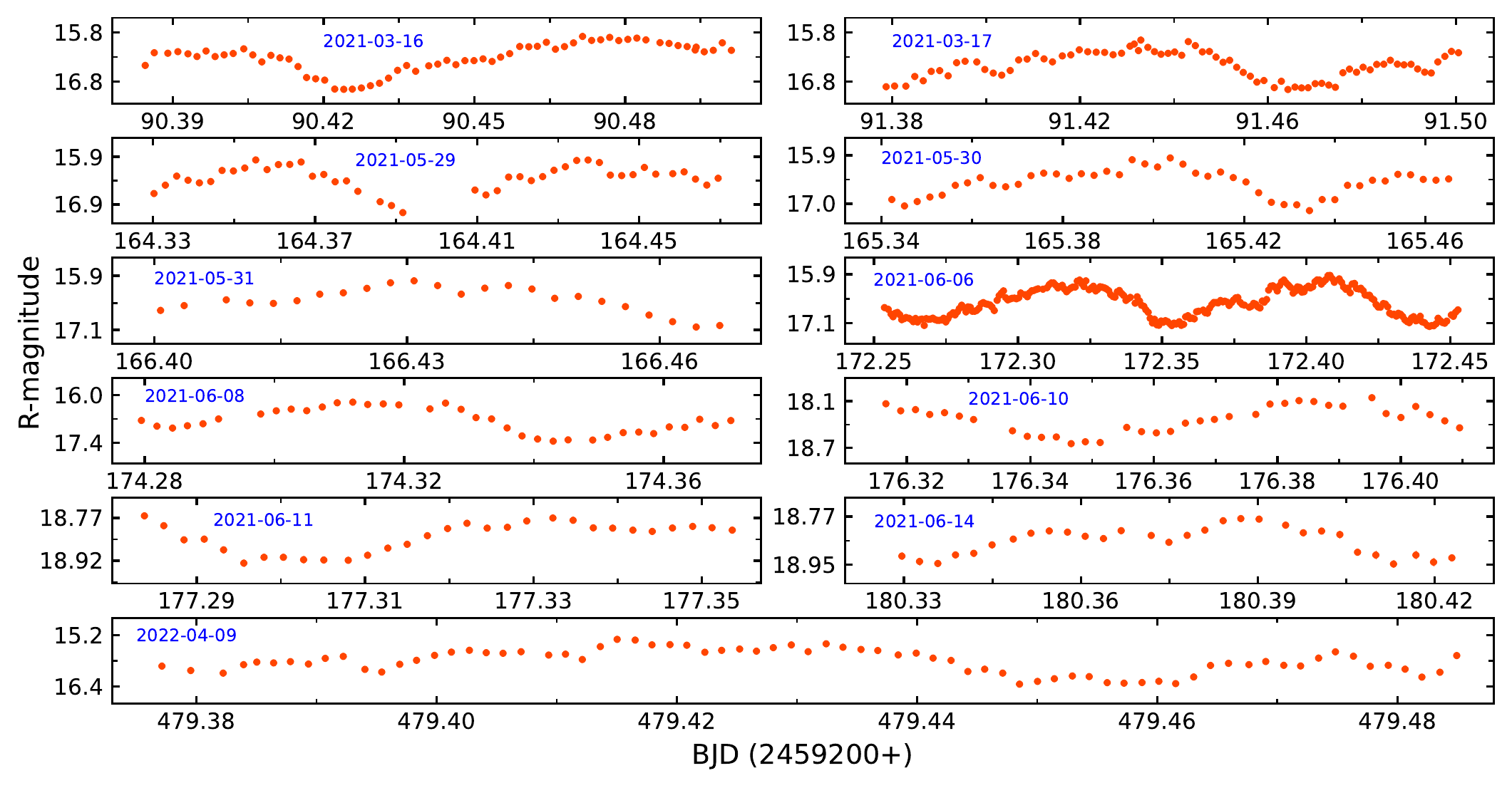} \label{fig:J1743_lc_opt}}
\subfigure[]{\includegraphics[width=0.82\textwidth]{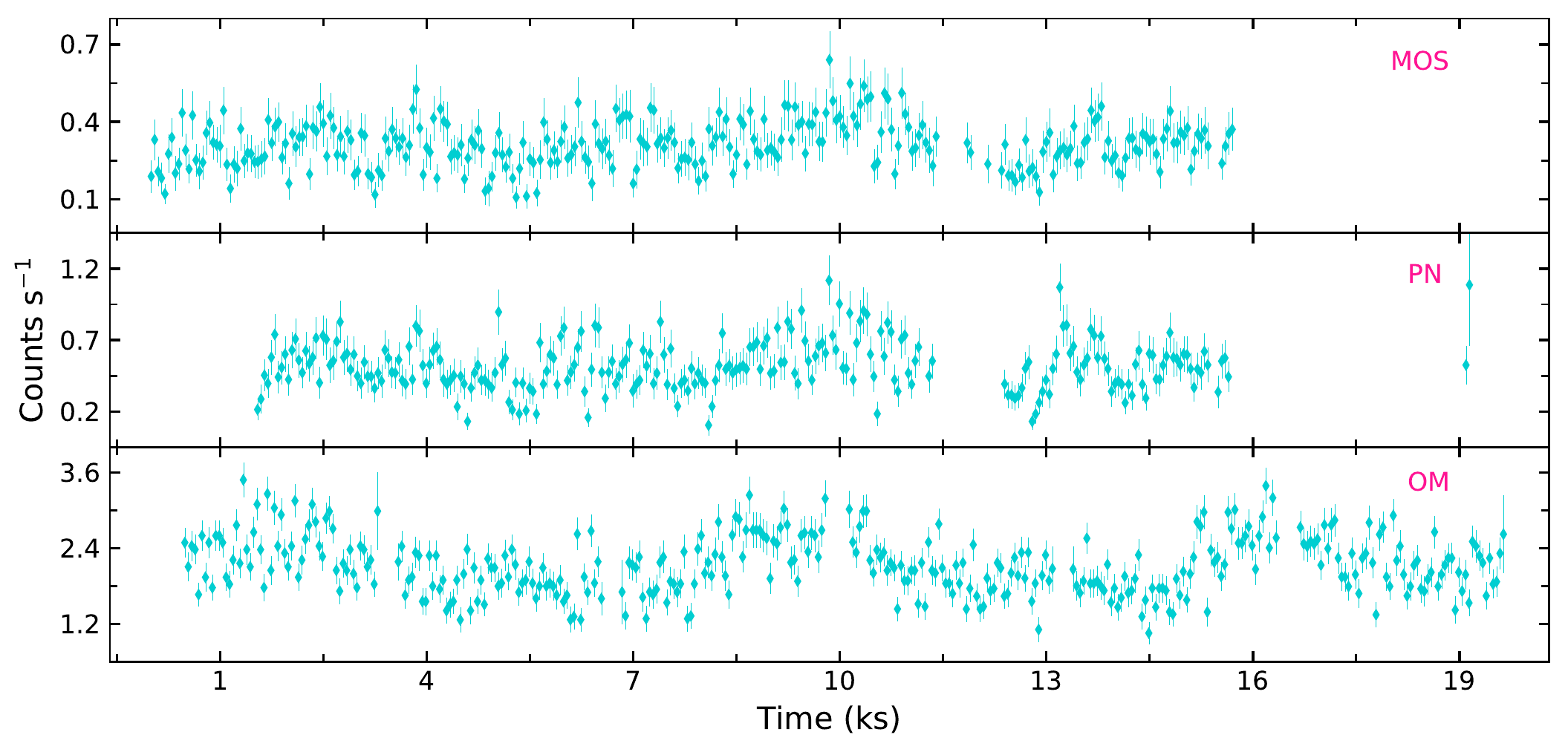} 
\label{fig:J1743_lc_xray}}
\caption{Optical, X-ray, and UV light curves of J1743. (a) The long-term ASAS-SN along with R-band observations. (b) Present R-band observations where the date of each observation is mentioned over each light curve. J1743 experienced a low-accretion state in June 2021, which can be seen in the light curves of three epochs: June 10, June 11, and June 14. (c) EPIC light curves in the 0.3-10.0 keV energy range and OM light curve in UVW1 filter binned in 50 s intervals.}
\label{fig:J1743_lc}
\end{figure*}

\subsection{ASAS-SN, \textit{TESS}, AAVSO, and CRTS observations} \label{sec2.2}
We have utilized the publicly available V-band data of J1743 from the All-Sky Automated Survey for Supernovae \citep[ASAS-SN;\footnote{\url{https://asas-sn.osu.edu/variables}}][]{2014ApJ...788...48S, 2017PASP..129j4502K} for our analysis. 
\par For YY Sex, we used archival data from the Transiting Exoplanet Satellite Survey (\textit{TESS}), the American Association of Variable Star Observers (AAVSO), and the Catalina Real-time Transient Survey (CRTS). The \textit{TESS} observations of YY Sex (TIC 62845887) were carried out from 2021 February 09 to 2021 March 06 at a cadence of 2 min. \textit{TESS} is equipped with four 10.5 cm telescopes that observe a  24\textdegree$\times$96\textdegree ~strip of sky, also known as sectors \citep[see][for details]{2015JATIS...1a4003R}. The total observing time was  $\sim$24 d, with a gap of  $\sim$4.9 d in the middle.  The data were extracted from the Mikulski Archive for Space Telescopes (MAST) data archive\footnote{\url{https://mast.stsci.edu/portal/Mashup/Clients/Mast/Portal.html}}. We have used the PDCSAP$\textunderscore$FLUX data values, which are the simple aperture photometry (SAP$\textunderscore$FLUX) values, after correction for systematic trends. We did not consider photometric points which did not have a `quality flag' value of 0. We have also utilized AAVSO\footnote{\url{https:// www.aavso.org/}} \citep{2021.K} CV (unfiltered data with a V-band zero-point), CR (unfiltered data with an R-band zero-point), V (Johnson V), and CRTS\footnote{\url{http://nunuku.caltech.edu/cgi-bin/getcssconedbid_release2.cgi}} \citep{2009ApJ...696..870D} V bands data to represent the long-term variability of the source.

% Figure-2
\begin{figure*}
\centering
\subfigure[]{
\includegraphics[width=0.45\textwidth]{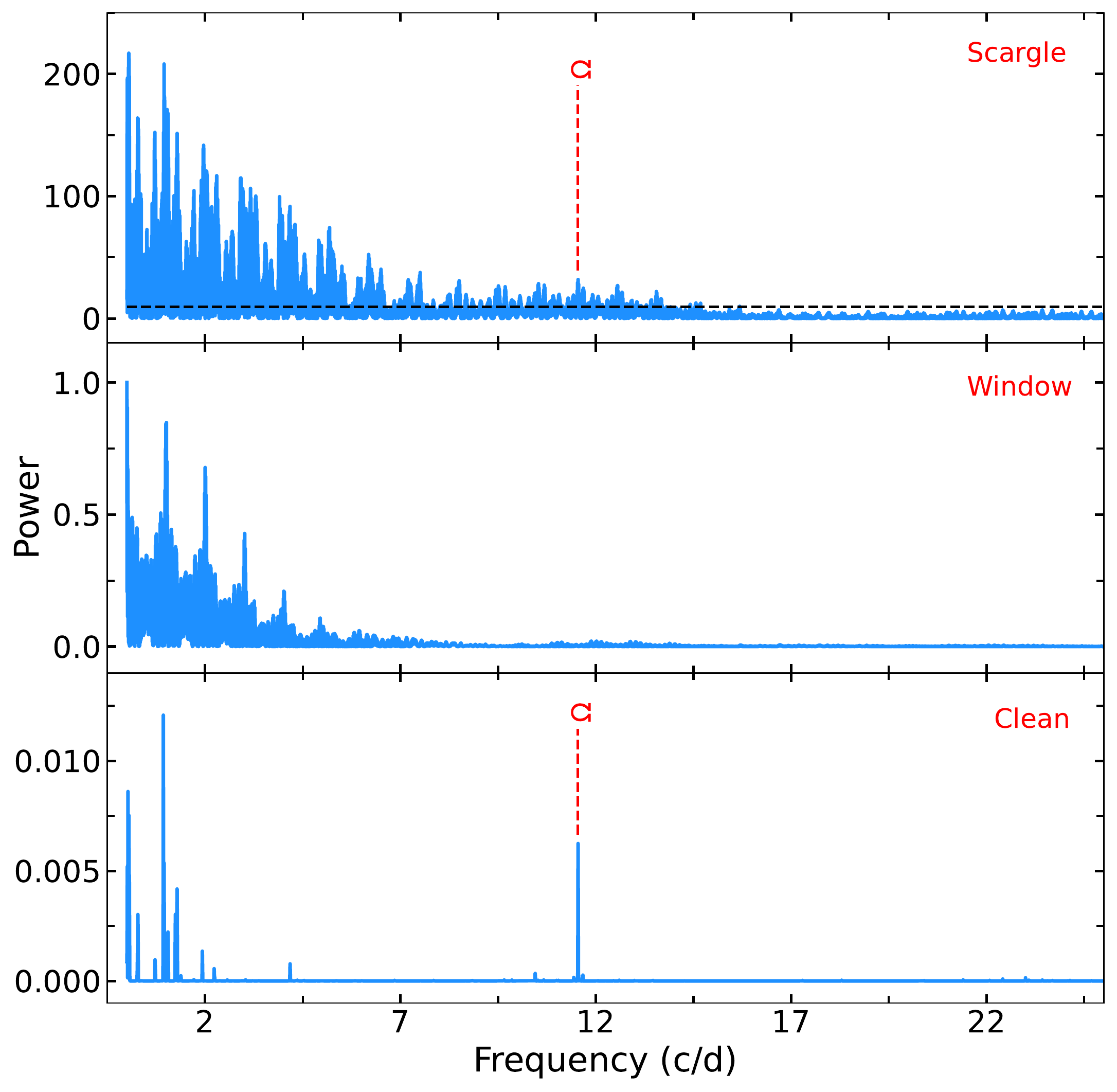}
\label{fig:J1743_opt_ps}}
\subfigure[]{\includegraphics[width=.45\textwidth]{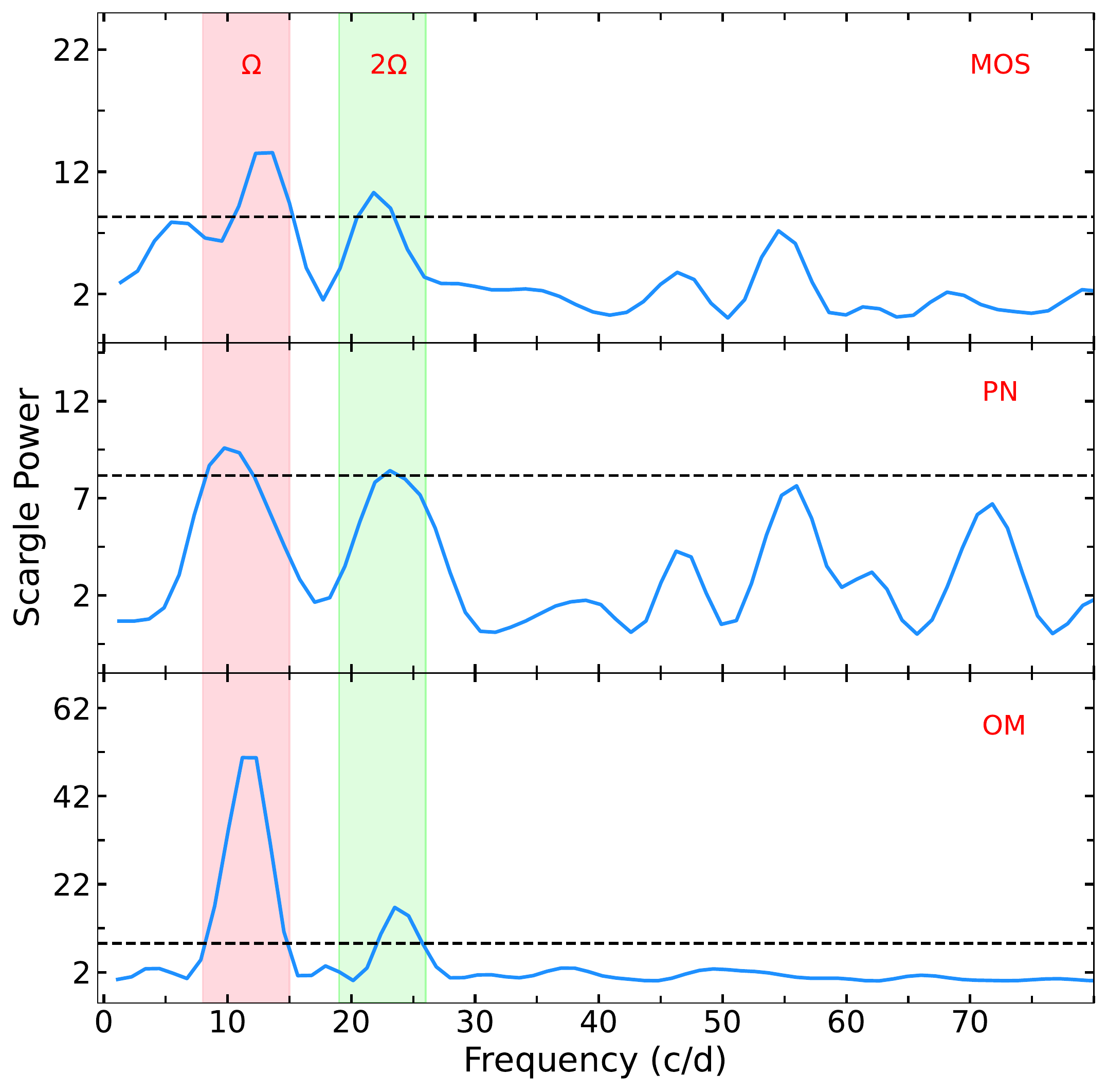}
\label{fig:J1743_xray_ps}}
\caption{(a) LSP (top panel), window function (middle panel), and CLEANed power spectra (bottom panel) of J1743 as obtained from ground-based optical observations. (b) The top and middle panels show the LSP of J1743 obtained from EPIC-MOS and EPIC-PN in the 0.3-10.0 energy band, whereas the bottom panel shows LSP obtained from OM observations. The shaded regions represent the \orb ~and \twoorb ~frequency regions, which are, in turn, frequency$\pm$error. In each plot, the horizontal dashed line represents 90 per cent significance level.}
\label{fig:J1743_ps}
\end{figure*}

%Table-3
\begin{table*}
\begin{center}
\caption{Periods corresponding to dominant peaks in the LSP of J1743 and YY Sex obtained from optical and X-ray data.}
\label{tab:periods}
\end{center}
\renewcommand{\arraystretch}{1.3}
\begin{tabular}{lccccccc}
\hline
Object & Period (h) & Optical-R & MOS & PN & OM & \textit{TESS} & AAVSO \\
\hline
J1743 & \po & 2.0784 $\pm$ 0.0001 & 1.76 $\pm$ 0.18 & 2.46 $\pm$ 0.31 & 2.14 $\pm$ 0.14 & .....& ..... \\
& \ptwoo & ..... & 1.10 $\pm$ 0.07 & 1.04 $\pm$ 0.05 & 1.02 $\pm$ 0.05 & ..... & ..... \\

\hline
YY Sex & \po & 1.5746 $\pm$ 0.0002 & ..... & ..... & ..... & 1.5746 $\pm$ 0.0011 & 1.574526 $\pm$ 0.000004\\
& \ptwoo & ..... & ..... & ..... & ..... & 0.7873 $\pm$ 0.0003 & .....\\
& \pthreeo & ..... & ..... & ..... & ..... & 0.5249 $\pm$ 0.0001 & .....\\
\hline
\end{tabular}
\end{table*}

\subsection{Optical spectroscopic observations} \label{sec2.3}
The spectroscopic observations of both sources were obtained using the 2-m Himalayan Chandra Telescope (HCT) at IAO, Hanle, equipped with the Hanle Faint Object Spectrograph and Camera (HFOSC). For our observations, we have used the grism Gr7 (3800-6840 \AA), which has a spectral resolution of 1330. The observing log for these observations is given in Table \ref{tab:obslog}. During each observing run, spectrophotometric standard stars and FeAr arc lamps were also observed  for flux calibration and wavelength calibration, respectively. The spectra were extracted using standard tasks in \textsc{iraf} and the reduced flux calibrated spectra were used for further analysis.

\subsection{X-ray observations } \label{sec2.4}
J1743 was also observed at X-ray wavelengths by the \textit{XMM-Newton} satellite  \citep{2001A&A...365L...1J} on 2019 September 21 using the European Photon Imaging Camera \citep[EPIC;][]{2001A&A...365L..18S, 2001A&A...365L..27T} with observation ID 0842570301. The total duration of observations was 21 ks. The standard \textit{XMM-Newton} \textsc{science analysis system (sas)} software package (version 20.0.0) with the latest calibration files\footnote{\url{https://www.cosmos.esa.int/web/XMM-Newton/current-calibration-files}} was used for the data reduction. We used SAS tools \textit{emproc} and \textit{epproc} to produce calibrated event files. We have inspected the data for the high background proton flares and removed the time corresponding to these flares. The \textit{epatplot} task was used to check the existence of pile-up, but we did not find any significant presence of it. The \textit{barycen} task was also used to correct the event for arrival times to the solar system barycenter.   Further analyses were carried out in the energy band of  0.3-10.0 keV. To extract the final light curve, spectrum, and detector response files, we have selected a circular region with a radius of 30 arcsec centring the source. The background was also selected from a circular region with a similar size to that of the source in the same CCD.  The spectra have been rebinned to a minimum of 20 counts per bin with the \textit{grppha} tool. Further, temporal and spectral analyses were performed using \textsc{heasoft} version 6.29. J1743 was also observed  in the UVW1 filter using the optical monitor \citep[OM;][]{2001A&A...365L..36M}. The UVW1 filter has an effective wavelength of 2910 \AA ~\citep{2004SPIE.5488..103K}. The task \textit{omfchain} was used to process the OM fast mode data. There were 6 observations, with each lasting about 2.95 ks. These individual exposures were then merged and the corresponding light curve file was used for further analysis. 
\par YY Sex was also observed by the \textit{XMM-Newton} satellite on two epochs, 2004 May 22 and 2004 June 21, for a total duration of 24 ks and 8.9 ks with observations IDs 0201290101 and 0201290401, respectively. For both IDs, the data were reduced using the same steps as discussed earlier. However, the source was not detected in the calibrated event files of observation ID 0201290401; therefore, we could not extract the light curve and spectrum. 
%light curve and spectrum extraction was not done. 

\subsection{Photo-polarimetric observations } \label{sec2.5}
Time-resolved photo-polarimetry using the two-channel all Stokes polarimeter, HIPPO \citep{Potter2010}, was conducted
on the SAAO 1.9-m telescope on 2010 March 12. Observations were taken with a broad-band red filter (OG57). Waveplate position angle offsets, instrumental polarization, and efficiency factors were calculated from observations of polarized and non-polarized standard stars during the observing run. Background sky measurements were taken frequently during the course of the observations. The photometry is not absolutely calibrated and instead is given as total counts minus the background-sky counts. Although the photometry and polarization data are accumulated every 0.01 and 0.1 s, respectively, the data were binned to 40 s and 120 s, respectively, in order to increase the signal-to-noise ratio.

% Figure-3
\begin{figure*}
\centering
\subfigure[]{
\includegraphics[width=0.85\textwidth, height=8cm]{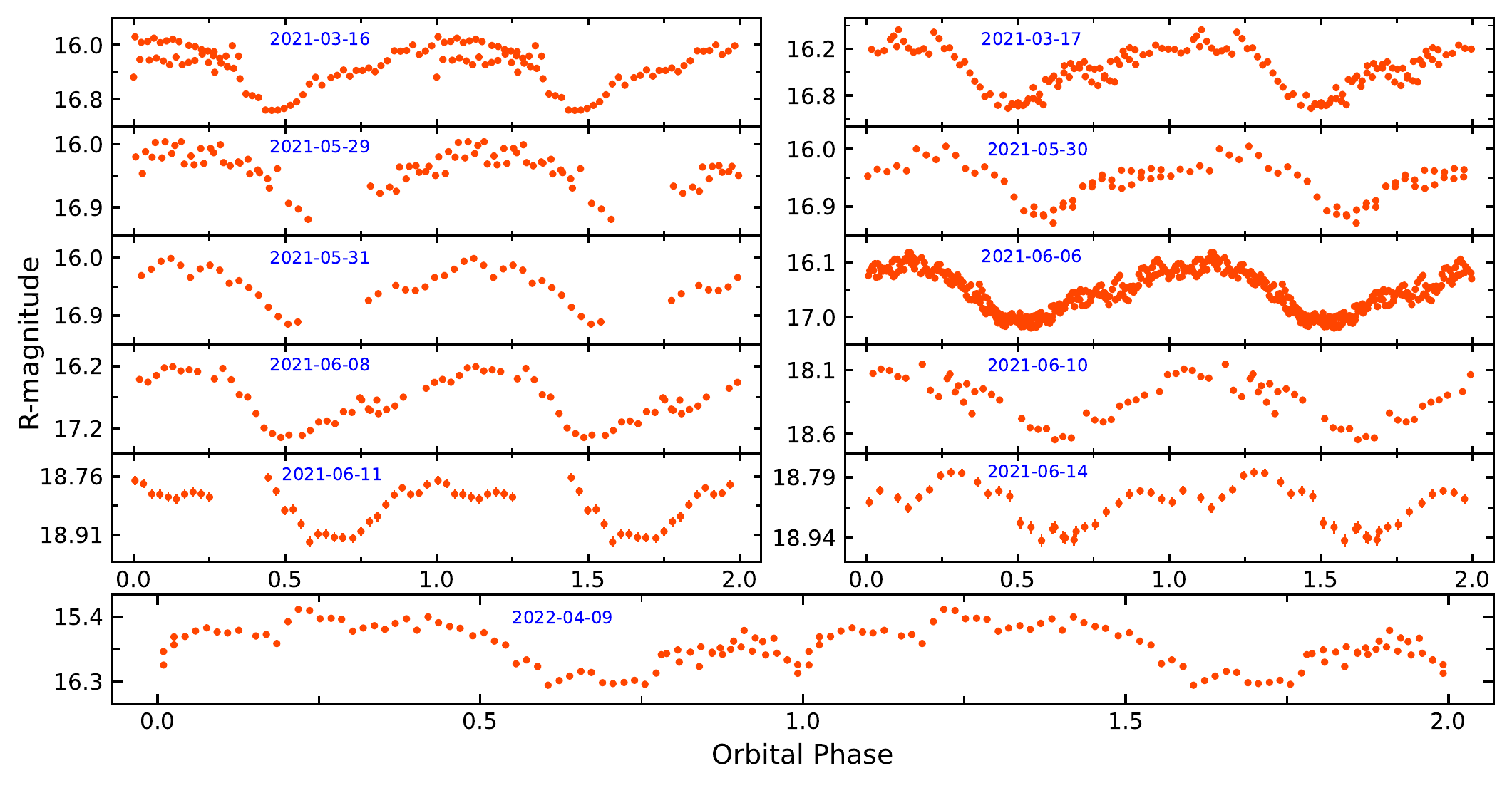}
\label{fig:J1743_opt_folded_alone}}
\subfigure[]{
\includegraphics[width=0.85\textwidth]{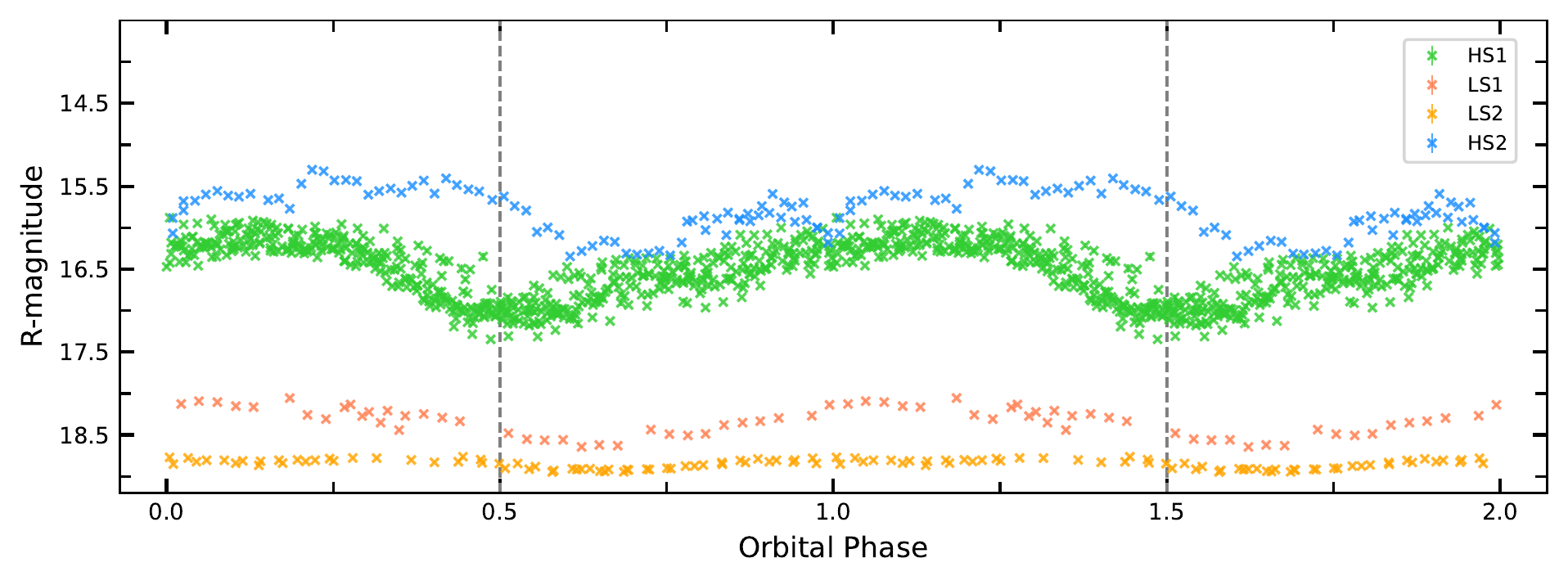}
\label{fig:J1743_opt_folded_all}}
\caption{Orbital phase folded R-band light curves of J1743. (a) Individual observations (b) High-state (HS1 and HS2) and low-state (LS1 and LS2) observations (Please see text for details).}
\label{fig:J1743_opt_folded}
\end{figure*}

% Figure-4
\begin{figure*}
\centering
\subfigure[]{\includegraphics[width=10cm, height=4cm]{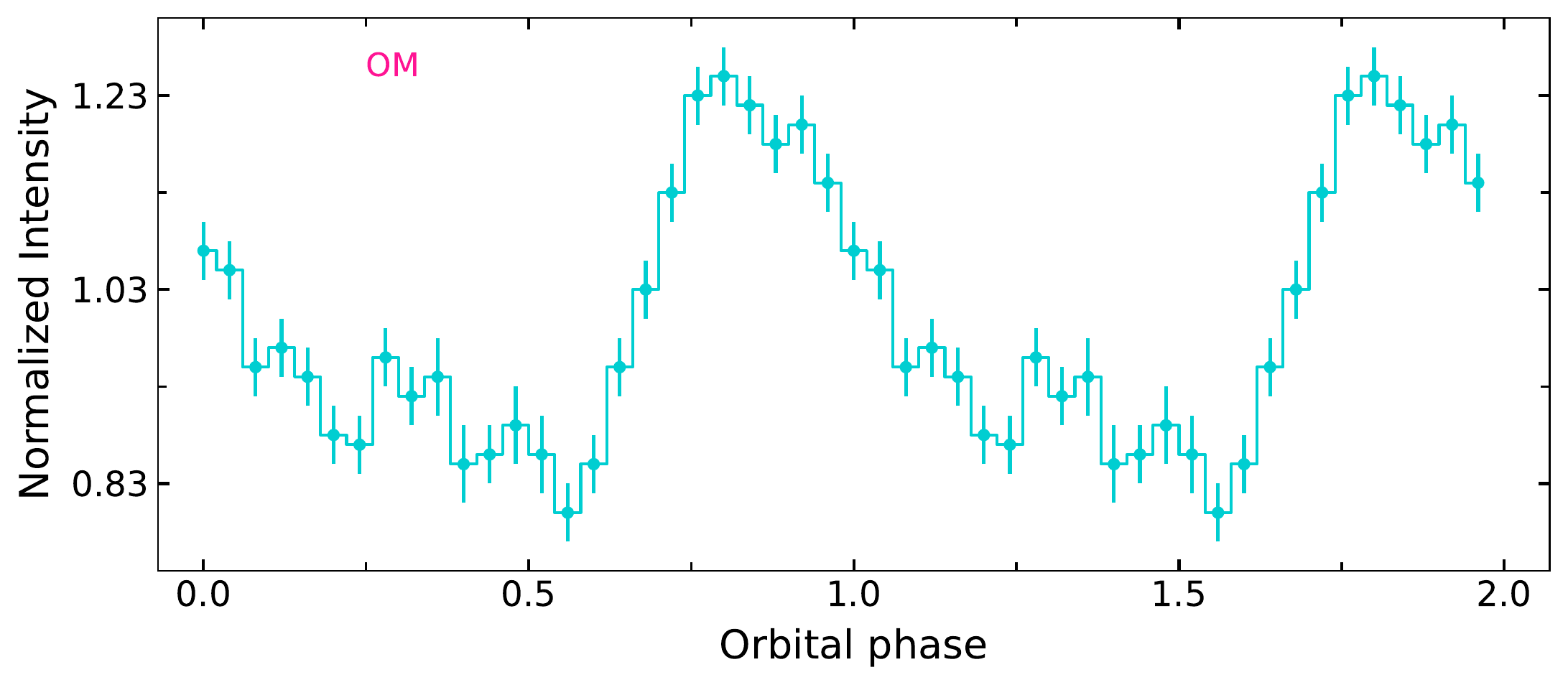} \label{fig:J1743_folded_om}}
\subfigure[]{\includegraphics[width=0.45\textwidth]{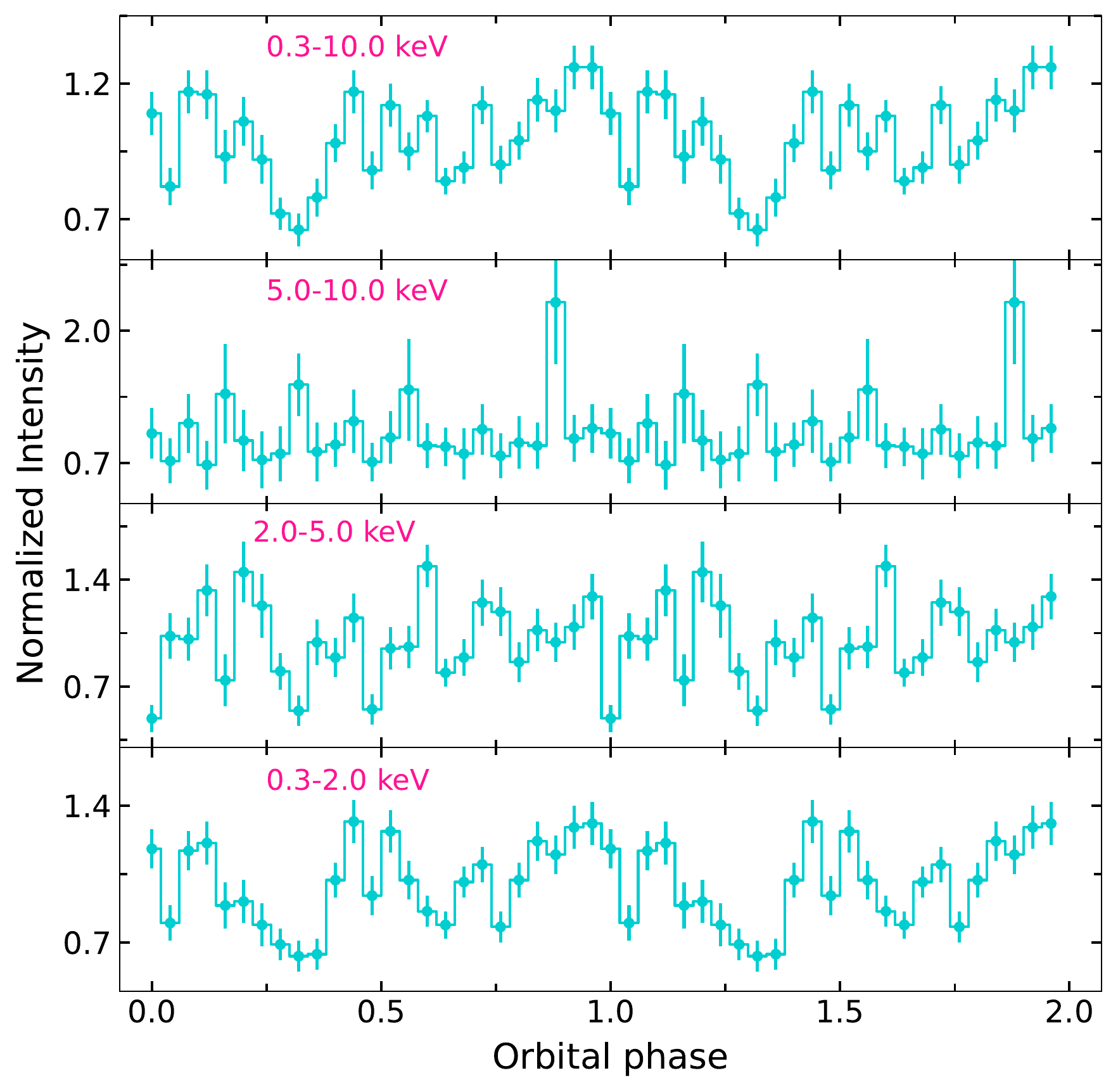}
\label{fig:J1743_folded_mos}}
\subfigure[]{\includegraphics[width=0.45\textwidth]{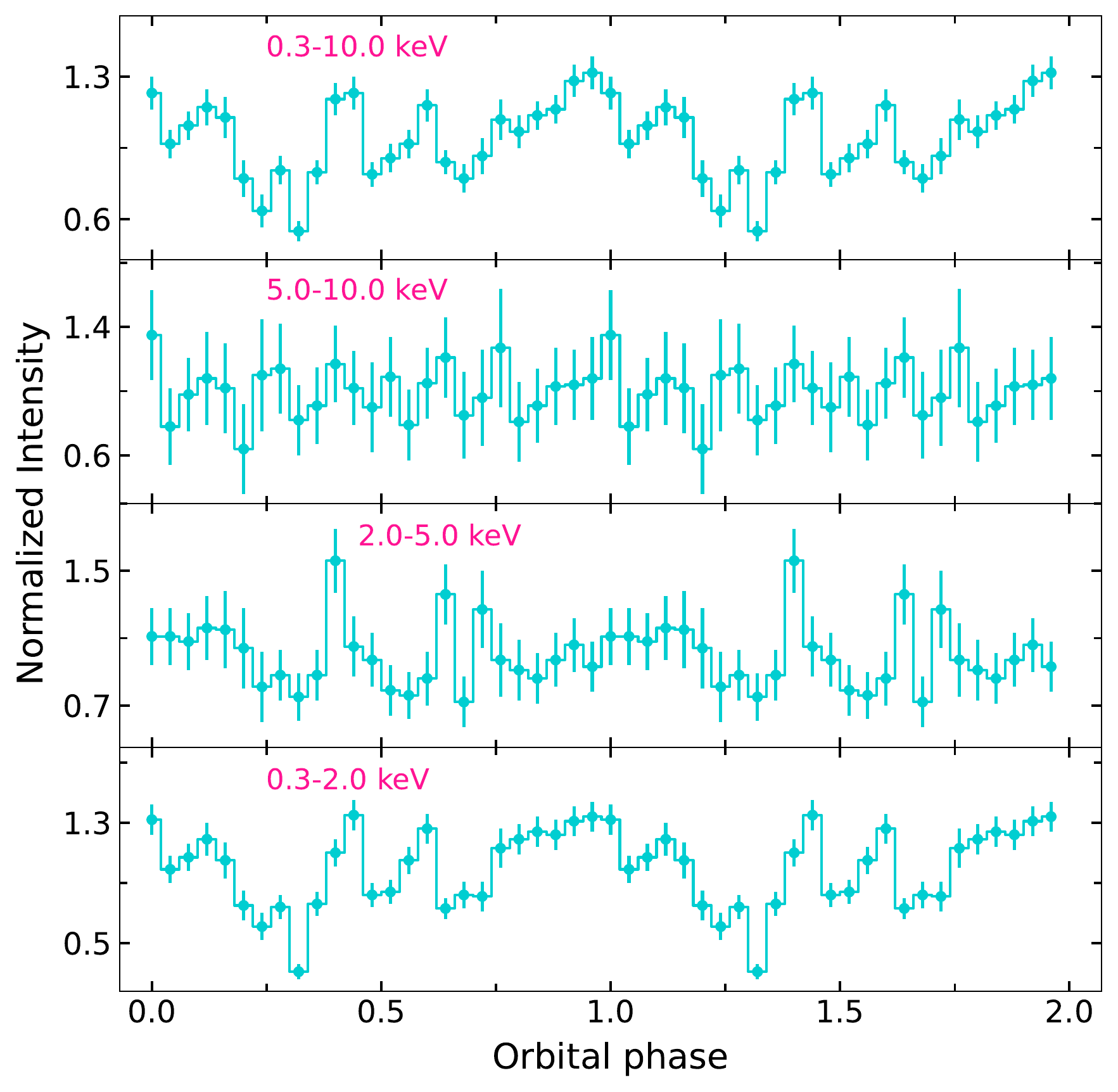} 
\label{fig:J1743_folded_pn}}

\caption{Orbital phase folded light curves observed from  (a) OM , (b) EPIC-MOS, and (c) EPIC-PN }
\label{fig:J1743_folded_xmm}
\end{figure*}

% Figure-5
\begin{figure*}
\centering
\includegraphics[width=0.8\textwidth]{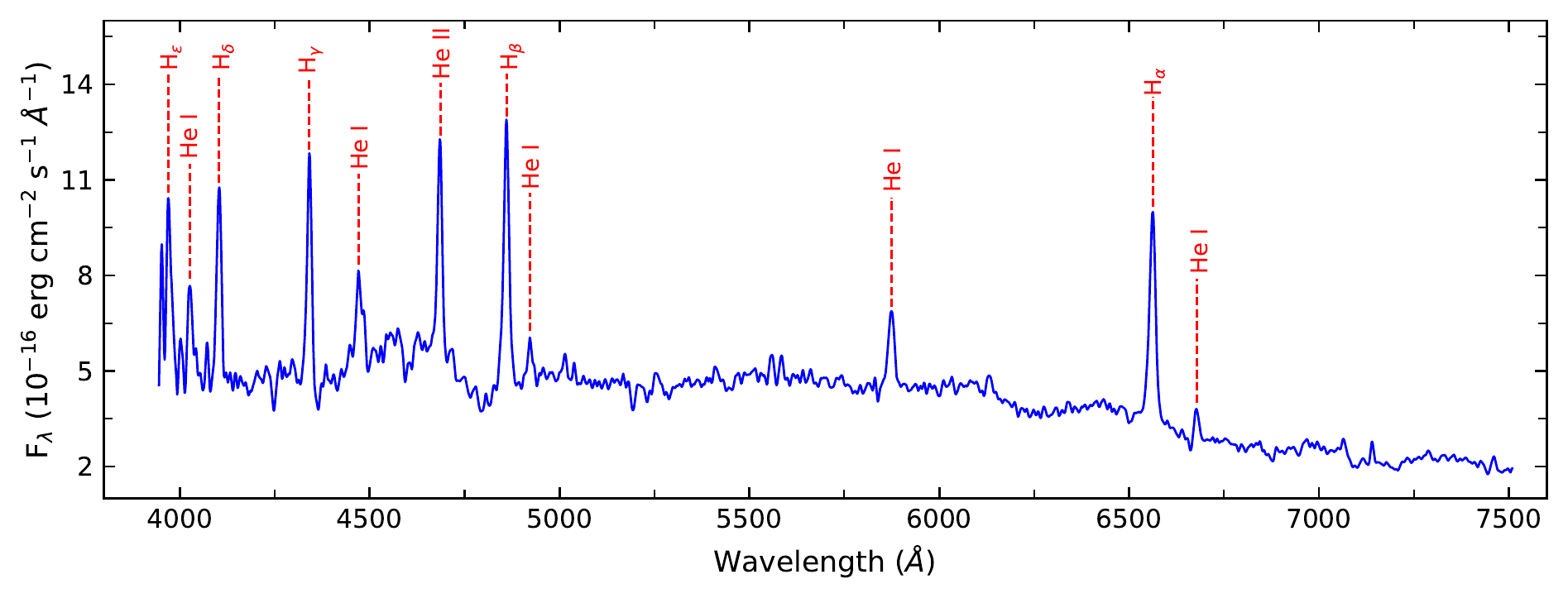}
\caption{Optical spectrum of J1743.}
\label{fig:J1743_spec_opt}
\end{figure*}

% Table-4
\begin{table*}
\renewcommand{\arraystretch}{1.3}
	\centering
\caption {Identiﬁcation, flux, EW, and FWHM for emission features in the spectra of J1743 and YY Sex.} \label{tab:opt_spec}
\begin{tabular}{lcccccccccc}
\hline
\multirow{3}{*}{Identification} && \multicolumn{3}{c}{J1743} && \multicolumn{3}{c}{YY Sex} \\
\cline{3-5} \cline{7-9} 
 && \multicolumn{3}{c}{2021 May 28} && \multicolumn{3}{c}{2022 Jan 24} \\
 \cline{3-5} \cline{7-9}
 && Flux & -EW & FWHM && Flux & -EW & FWHM \\
 \hline
H $\eta$ (3835 \AA) && ... & ... & ... && 2.16 & 9 & 1170\\
H $\zeta$ (3889 \AA) && ... & ... & ... &&  4.00 & 19 & 1172\\
H $\epsilon$ (3970 \AA) && 8.49 & 17 & 1189  && 4.71 & 24 & 1328 \\
He \Romannum{1} (4026 \AA) && 5.56 & 13 & 1198 && 1.90 & 11 & 1183 \\
H $\delta$ (4102 \AA) && 9.54 & 14 & 1049 &&  7.00 & 36 & 1781 \\ 
H $\gamma$ (4340 \AA) && 11.03 & 25 & 991 &&  5.51 & 26 & 1104\\
He \Romannum{1} (4471 \AA) && 5.91 & 11 & 1486  && 2.66 & 15 & 1342 \\
He \Romannum{2} (4686 \AA) && 10.21 & 19 & 907 && 4.70 & 25 & 1185\\
H $\beta$ (4861 \AA) && 14.75 & 33 & 1058 && 8.62 & 52 & 1393\\
He \Romannum{1} (4922 \AA) && 3.02 & 7 & 1236 &&  0.88 & 5 & 1016 \\
He \Romannum{1} (5875 \AA) && 5.32 & 12 & 1022 && 2.10 & 14 & 1208 \\
H $\alpha$ (6563 \AA) && 12.69 & 36 & 861 &&  6.11 & 42 & 1121 \\
He \Romannum{1} (6678 \AA) && 1.91 & 7 & 679 &&  1.08 & 8 & 894 \\
\hline
\end{tabular}

\bigskip
\emph{\textbf{Note.} Flux, EW, and FWHM are in the unit of $10^{-15}$ $\rm erg ~cm^{-2}$ $\rm s^{-1}$, \AA, and $\rm km ~s^{-1}$, respectively.}
\end{table*}

% Figure-6
\begin{figure*}
\centering
\includegraphics[width=0.7\textwidth]{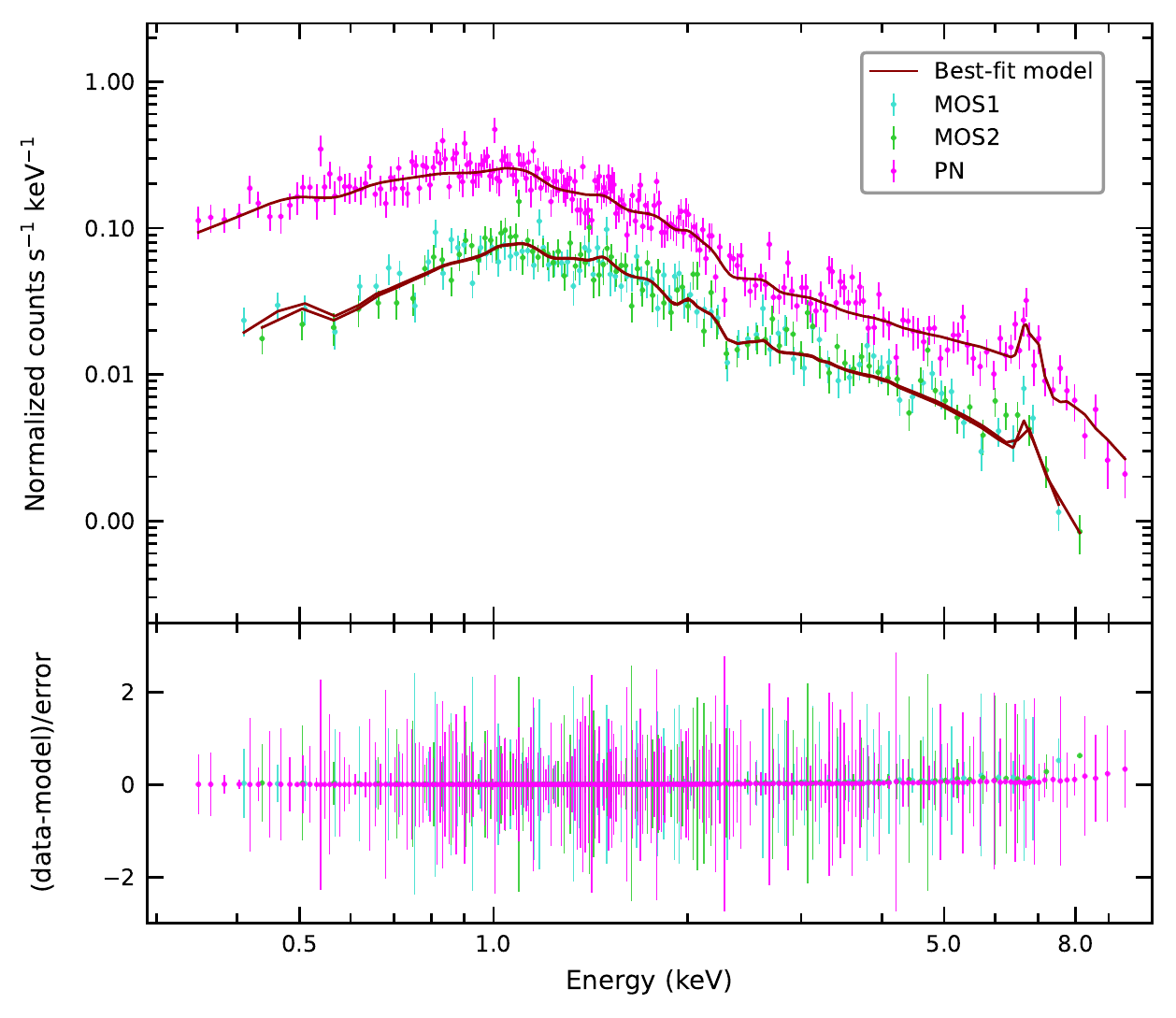}
\caption{X-ray spectra of J1743 along with the best-fitting model. The bottom panel shows the $\chi_\nu^2$ contribution of data points for the best-fitting model in terms of residual.}
\label{fig:J1743_spec_xray}
\end{figure*}

% Table-5
\begin{table*}
\renewcommand{\arraystretch}{1.3}
	\centering
\caption {Spectral parameters as obtained from the best-fitting model `phabs$\times$pcfabs$\times$(bb+mkcflow)' to the X-ray spectra of J1743. All the errors quoted here are with 90 per cent confidence range of a single parameter.} \label{tab:avg_spec}
\begin{tabular}{p{0.35\columnwidth}p{0.4\columnwidth}p{0.25\columnwidth}}
\hline
Model & Parameters  &  Value \\
\hline
\multirow{2}{*}{pcfabs}& N$_{\rm H,pcf}$ ($10^{23} \rm ~cm^{-2}$)  & $7.5_{-5.5}^{+6.6}$\\
& pcf( per cent)  & $56_{-19}^{+20} $  \\
\multirow{2}{*}{bb}& T$_{\rm bb}$ (eV) &  $97_{-51}^{+52} $   \\
& N$_{\rm bb}$ (10$^{-6}$)    &  $3.6_{-1.6}^{+20.6} $  \\
\multirow{3}{*}{mkcflow} & T$_{\rm low}$ (keV)  &  < 0.84  \\
& T$_{\rm high}$ (keV) & $31.9_{-8.9}^{+12.9} $  \\
& N$_{\rm mkcflow}$ (10$^{-12}$ M$\odot$ yr$^{-1}$)   &  $5.1_{-1.6}^{+4.2} $ \\
\multirow{1}{*}{Soft flux} & F$_{\rm s}$ (10$^{-13}$ erg cm$^{-2}$ s$^{-1}$)    &    $3.3_{-0.9}^{+0.9}$  \\
\multirow{1}{*}{Hard flux} & F$_{\rm h}$ (10$^{-12}$ erg cm$^{-2}$ s$^{-1}$)    &    $7.3_{-0.1}^{+0.1}$   \\
\multirow{1}{*}{Bolometric flux} & F$_{\rm bol}$ (10$^{-12}$ erg cm$^{-2}$ s$^{-1}$)    &    $7.6_{-0.1}^{+0.1}$   \\
\multirow{1}{*}{Bolometric luminosity} & L$_{\rm bol}$ (10$^{31}$ erg s$^{-1}$)    &    $4.21_{-0.08}^{+0.08}$   \\
& $\chi_\nu^2$ (dof)                       &  1.07(404)           \\
\hline
\end{tabular}

\bigskip
\emph{\textbf{Note.}} N$_{\rm H,pcf}$ is the partial covering absorber density i.e., absorption due to the partial covering of the X-ray source by the neutral hydrogen column; pcf is the covering fraction of the partial absorber; T$_{\rm bb}$ and N$_{\rm bb}$ are the blackbody temperature and the normalization constant of bb; T$_{\rm low}$ and T$_{\rm high}$ are the low and the high temperatures of mkcflow; N$_{\rm mkcflow}$ is the normalization constant of mkcflow; F$_{\rm s}$ and F$_{\rm h}$ are the soft and hard flux for soft (bb) and hard (mkcflow) components in the 0.001-100.0 keV energy band; F$_{\rm bol}$ is the unabsorbed bolometric flux derived for 0.001-100.0 keV energy band; L$_{\rm bol}$ is the corresponding bolometric luminosity calculated by assuming a distance of 214 pc.
\end{table*}

\section{Analysis and Results} \label{sec3}
\subsection{1RXS J174320.1-042953} \label{sec3.1}
\subsubsection{Optical and X-ray photometry} \label{sec3.1.1}
Fig. \ref{fig:J1743_lc} shows the light curves of J1743 in the optical and X-ray bands during different observing runs and Fig. \ref{fig:J1743_lc_all} displays the long-term variable nature of the source using ASAS-SN-V, OM-UVW1, and our R-band observations. J1743 displays a highly variable light curve at optical wavelengths. The  R-band light curves, as shown in Fig. \ref{fig:J1743_lc_opt}, show  a change in the system's state at three epochs, i.e. 2021 June 10, 11, and 14. The sudden change in brightness indicates that J1743 entered into a low state on June 10, which we refer to as LS1. While the latter two low-states  observations are referred to as LS2. The average magnitude in these low states is in the range of  $\sim$18.0-18.95 mag. Almost a year later, J1743 was observed in the high state with a magnitude range of 15.3-16.3 mag, referred to as HS2. On the other hand, the high states  before entering into low states (referred to as HS1) were in the magnitude range of 15.88-17.3 mag. This indicates that the HS2 observations  occurred in a comparatively higher state than 2021's HS1 observations.  Two periodic cycles are clearly visible in the  longest duration ($\sim$4.8 h)  observations on 2021 June 06. The background-subtracted X-ray light curves of J1743 in the 0.3-10.0 keV energy range are shown in the top two panels of Fig. \ref{fig:J1743_lc_xray}, whereas the OM-UVW1 light curve is shown in the bottom panel of Fig. \ref{fig:J1743_lc_xray}. A temporal binning of 50 s was used for the extraction of both EPIC and OM light curves. The variable nature of the source is also evident in both X-ray and UV light curves. 

%The OM-UVW1 light curve shows almost 3 cycles of these variability patterns.

\par The periodicity present in the data was extracted by employing the  Lomb-Scargle periodogram (LSP) method  \citep{1976Ap&SS..39..447L,1982ApJ...263..835S}, where an error in the peak is derived by calculating the half-size of a single frequency bin, centred on the peak, and then converting it to period units.
%\textbf{We have used \textsc{Starlink's} \citep{2014ASPC..485..391C} \textit{period} for periodogram analysis, where an error in the peak is derived by calculating the half-size of a single frequency bin centred on the peak, and then converting it to period units. }
The LSP obtained from R-band  observations is shown in the top panel of Fig. \ref{fig:J1743_opt_ps}. The black-dashed horizontal line represents the 90 per cent significance level, which is calculated by using the algorithm of \citet{1986ApJ...302..757H}. The LSP of J1743 is noisy due to data gaps. Therefore, we have computed the window function with the same time sampling but a constant magnitude to represent the peaks caused by irregular data sampling, which is shown in the middle panel of Fig. \ref{fig:J1743_opt_ps}. Low frequencies present in the LSP appear to be due to these irregular sampling. Therefore, we have also implemented the CLEAN algorithm  \citep{1987AJ.....93..968R} with a loop gain of 0.1 and 1000 iterations to ensure the periodicity in the data. The CLEANed power spectrum is shown in the bottom panel of Fig. \ref{fig:J1743_opt_ps}. A clear peak  at  $\sim$11.55 c/d is seen in the CLEANed power spectrum, which is also visible in the LSP. The period derived from LSP corresponding to this peak comes out to be 2.0784 $\pm$ 0.0001 h, which is consistent with the orbital period reported by \cite{2012PZ.....32....3D}. The LSP of MOS, PN, and OM data are shown in Fig. \ref{fig:J1743_xray_ps}, where two prominent peaks are seen, which correspond to \po and \ptwoo in each power spectra.  The significant periods derived from optical-R, MOS, PN, and OM-UVW1 data are given in Table \ref{tab:periods}. 
%Due to the short duration of \textit{XMM-Newton} observations than optical observations, the error derived in the period is large.

\par We have explored the evolution of the orbital phased light curve of J1743 using the derived value of the orbital period. The reference time for folding was taken at the first point of optical observations, i.e. BJD 2459290.3845311. The orbital folded R-band light curves are shown in Fig. \ref{fig:J1743_opt_folded_alone}. All light curves during observations corresponding to HS1 appear to have a saw-tooth-shaped profile with a steep rise and a rapid decline. Similar profiles have also been seen in some of the observations of MT Dra, which is a two-pole accretor polar \citep{2002A&A...392..505S}. All of these light curves show broad maxima with a phase duration of  $\sim$1.15-1.25 and flat minima centring near phase  $\sim$0.5. A double-hump structure can be seen in the folded light curves of LS2, where the shape of the light curves is almost the same. The  light curves of  LS1  differ from that of LS2 in terms of the absence of a double-hump structure. The minima in these LS observations have a shift from HS1 observations and occur near phase 0.6. A drastic change can be seen in the folded light curve of HS2 observations, where the orbital minimum is shifted to a phase value of  $\sim$0.65 in comparison to the HS1 observations (see Fig. \ref{fig:J1743_opt_folded_all}). A similar phase shift during two different high states was also observed in polar RX J0203.8+2959 \citep{1998A&A...338..465S}. These changes can be explained by changes in the shape, size and (or) the location of the accretion region, which in turn is related to the change in the accretion rate. On the other hand, OM-UVW1 folded light curve is shown in Fig. \ref{fig:J1743_folded_om}, where two humps are clearly visible. These two humps differ very much in brightness. The prominent hump has a maximum at phase 0.8, whereas the second hump has a maximum at phase 0.3 with a minimum at phase $\sim$0.55.

\par We have also generated energy-dependent orbital phase folded light curves in the 0.3-2.0 keV, 2.0-5.0 keV, and 5.0-10.0 keV energy bands. These are shown in Figs \ref{fig:J1743_folded_mos} and \ref{fig:J1743_folded_pn}, and are very different compared to OM-UV.
A clear modulation with a broad and a narrow maximum is seen only in the soft (0.3-2.0 keV) energy band, whereas the  modulation appears to be absent in the harder energy bands (2.0-5.0 keV and 5.0-10.0 keV). The presence of modulation in the soft energy band is associated with the photoelectric absorption in the accretion stream. In comparison to OM-UV, broad and narrow maxima occur at phase  $\sim$0.9 and  $\sim$0.5, respectively, with a small dip near 0.7 and minima at  $\sim$0.3.  

% Figure-7
\begin{figure*}
\centering
\subfigure[]{\includegraphics[width=0.86\textwidth]{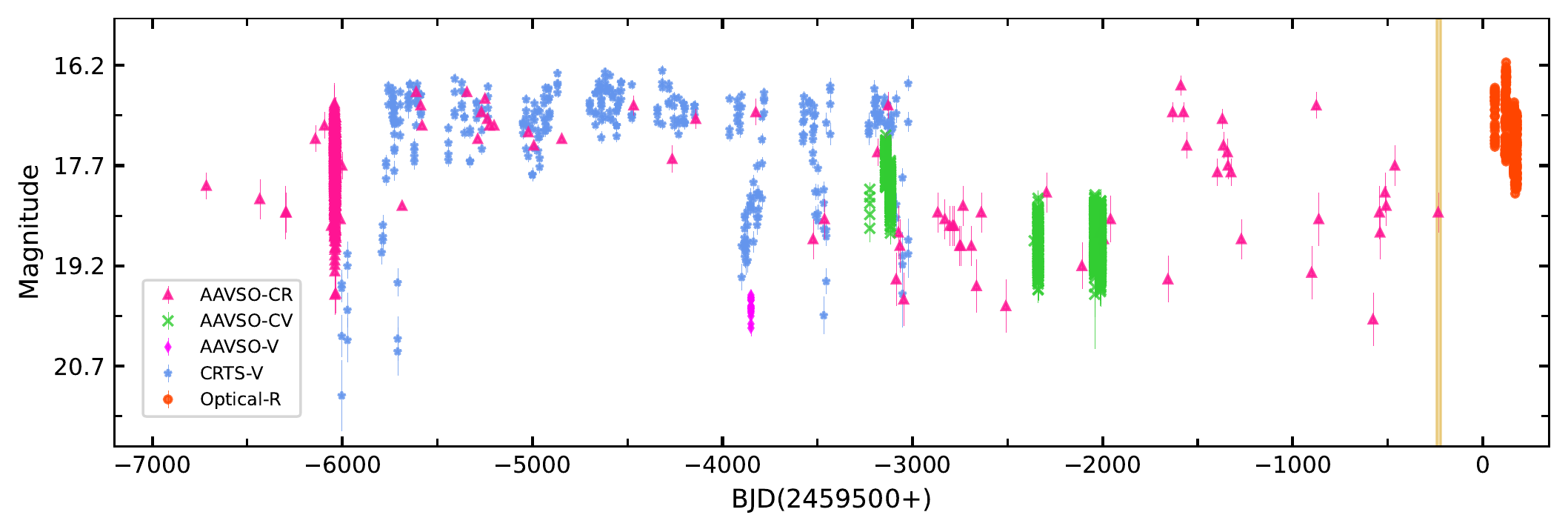} \label{fig:J1039_lc_all}}
\vskip -6pt
\subfigure[]{\includegraphics[width=0.86\textwidth]{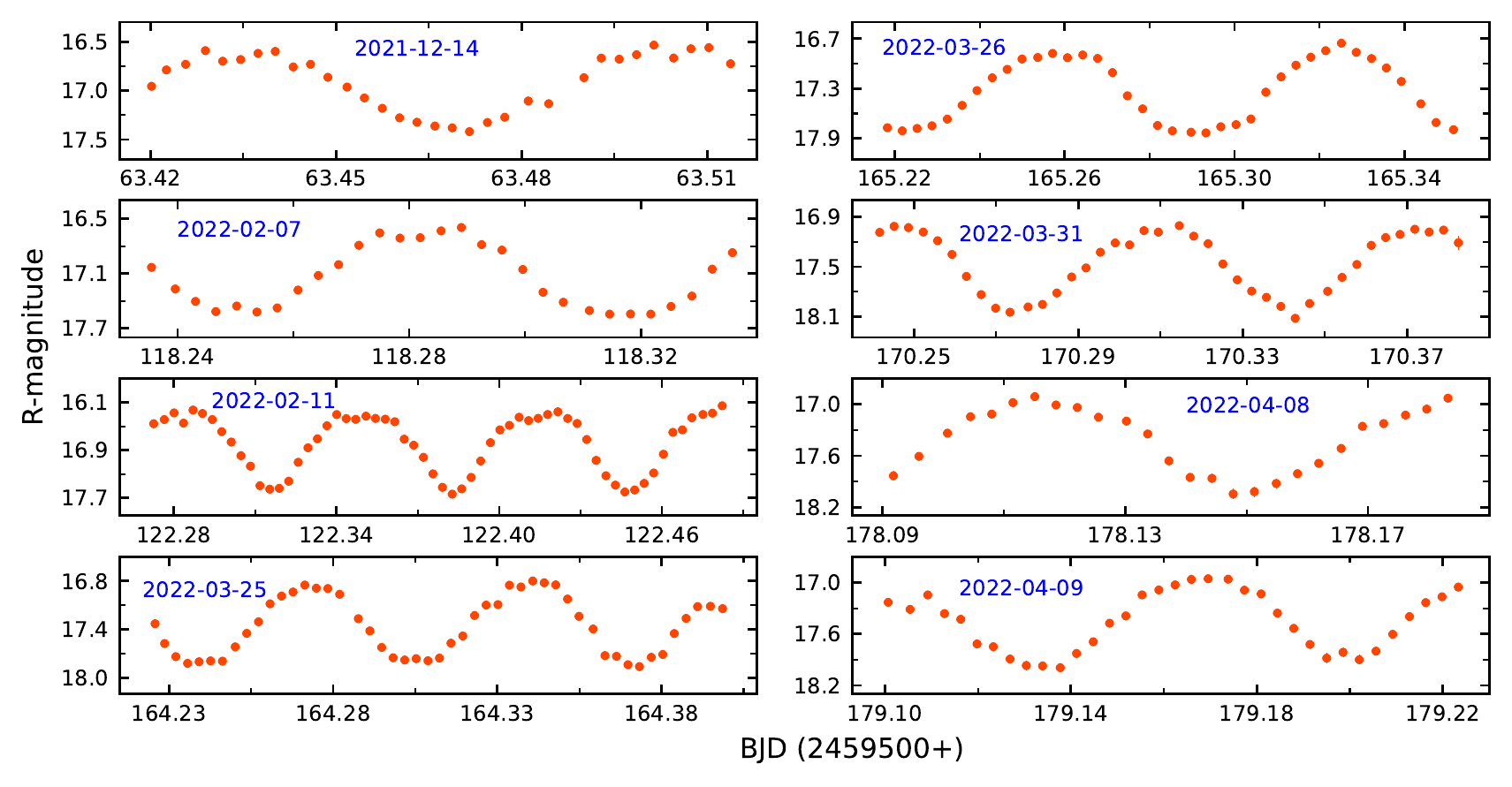} 
\label{fig:J1039_lc_opt}}
\vskip -6pt
\subfigure[]{\includegraphics[width=0.86\textwidth]{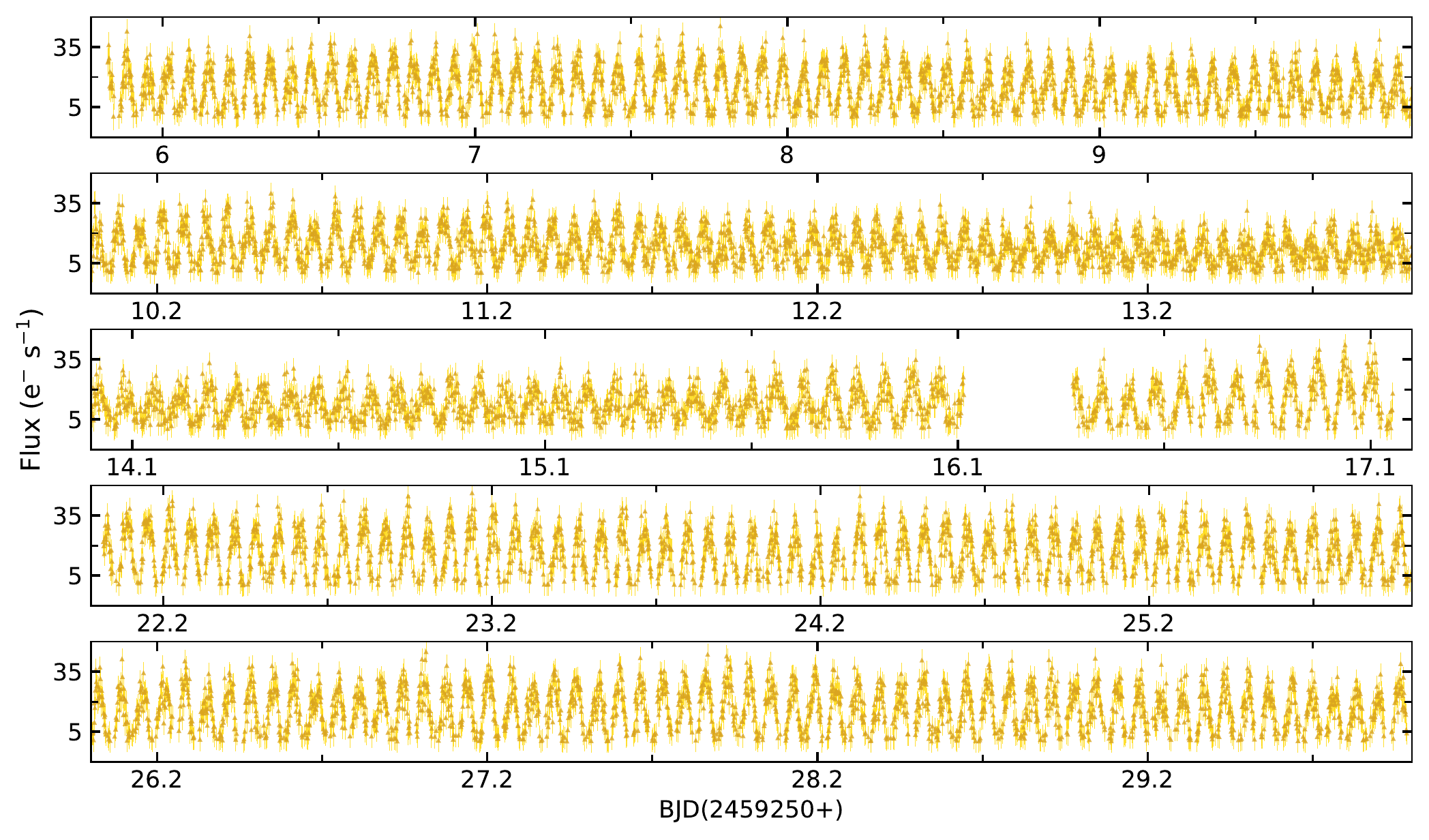} \label{fig:J1039_lc_tess}}
\vskip -6pt
\caption{The combined long-term AAVSO, CRTS, and our optical light curve of YY Sex. The time span of the \textit{TESS} observations is indicated in yellow, (b) R-band light curves of YY Sex where the corresponding dates of observations are mentioned near each light curve, and (c) \textit{TESS} light curve of YY Sex. }
\label{fig:J1039_lc}
\end{figure*}

\begin{comment}
% Figure-8
\begin{figure*}
\centering
\includegraphics[width=16cm, height=10cm]{J1039_tess_lc_revised.pdf}
\caption{\textit{TESS} light curve of YY Sex, where the flux is in units of e$^{-}$ s$^{-1}$.}
\label{fig:J1039_lc_tess}
\end{figure*}
\end{comment}

\subsubsection{Optical and X-ray spectroscopy} \label{sec3.1.2}
Fig. \ref{fig:J1743_spec_opt} shows the optical spectrum of J1743 with a wealth of spectral features. It exhibits strong single-peaked emission lines of H-Balmer (H$\alpha$ to H$\epsilon$) along with neutral and ionized He lines (He \rn{1} and He \rn{2}). Table \ref{tab:opt_spec} shows the identification, fluxes, equivalent width (EW), and FWHM of these lines. The Balmer lines appear with an inverted Balmer-decrement of F(H$\alpha$)/F(H$\beta$)=0.86. The He \rn{2} 4686 \AA ~is quite strong, which suggests the magnetic nature of accretion flow. The EW ratio of He \rn{2} 4686 \AA ~to {H$\beta$} comes out to be  $\sim$0.6. The observed spectrum resembles the spectrum of \citet{2017AJ....153..144O} observed in 2012. The values of flux and FWHM are found to be variable during the present and previous epochs of observation (see \citet{2017AJ....153..144O} and present work).

\par To perform the X-ray spectral analysis in the energy range 0.3-10.0 keV , we have used \textsc{xspec} version-12.12.0 \citep{1996ASPC..101...17A, 2001ASPC..238..415D}.  In general, the X-ray spectrum of polars is explained by a soft blackbody-like emission component superimposed on a multi-temperature thermal plasma emission along with an absorption component. Therefore, we used blackbody (bb) and  cooling-flow plasma emission model \citep[\texttt{mkcflow};][]{1988ASIC..229...53M}. Further, to account for the interstellar absorption \texttt{phabs} component was utilized. We have taken the abundance tables and the photoelectric absorption cross section `bcmc' from  \citet{2009ARA&A..47..481A} and \citet{1992ApJ...400..699B}, respectively. 
%The majority of the X-ray spectra of MCVs suffer from the local absorbers; therefore, to account for the local absorption effect in the spectral fitting, a partial covering absorption component \texttt{pcfabs} was used. 
The majority of the X-ray spectra of MCVs are affected by local absorbers; thus, a partial covering absorption component \texttt{pcfabs} was used to account for the local absorption effect in spectral fitting. The value of redshift required in the \texttt{mkcflow} model cannot be zero, therefore it was fixed to a value of 5 $\times$ 10$^{-8}$ for a cosmological Hubble constant of 70 km s$^{-1}$ $\rm Mpc^{-1}$ and a distance of 214$\pm$2 pc \citep{2021AJ....161..147B}. We used a value of 2 for the switch parameter, which determines the spectrum to be computed using AtomDB data.  We fixed the $\rm N_{H}$ value to the total Galactic column in the direction of J1743 of $1.79 \times 10^{21}$ cm$^{-2}$ \citep{2005A&A...440..775K}. In \textsc{xspec} notations, the adopted model to fit the X-ray spectra was implemented as  \texttt{phabs$\times$pcfabs$\times$(bb+mkcflow)}, which resulted in a remarkably good representation of data with a reduced chi-square ($\chi^{2}_{\nu}$) value of 1.07.  By incorporating the \texttt{cflux} model in the best-fitting model, we have calculated the unabsorbed soft (bb), hard (mkcflow), and bolometric flux in the 0.001-100 keV energy band. The  background-subtracted EPIC PN and MOS spectra of J1743, along with the best-fitting model, are shown in Fig. \ref{fig:J1743_spec_xray}. The best-fit spectral parameters estimated from the simultaneous fitting of both EPIC PN and MOS spectra are given in Table \ref{tab:avg_spec}. The X-ray spectrum of J1743 is well explained by a multi-temperature component with a high temperature of  $\sim$32 keV and a low temperature of < 840 eV. The derived value of the mass accretion rate from the spectral fitting is found to be  $\sim$5.1 $\times$ $10^{-12}$ M$\odot$ yr$^{-1}$. A blackbody component was found to be appropriate to explain the soft emission.

% Figure-8
\begin{figure*}
\centering
\subfigure[]{\includegraphics[width=0.45\textwidth]{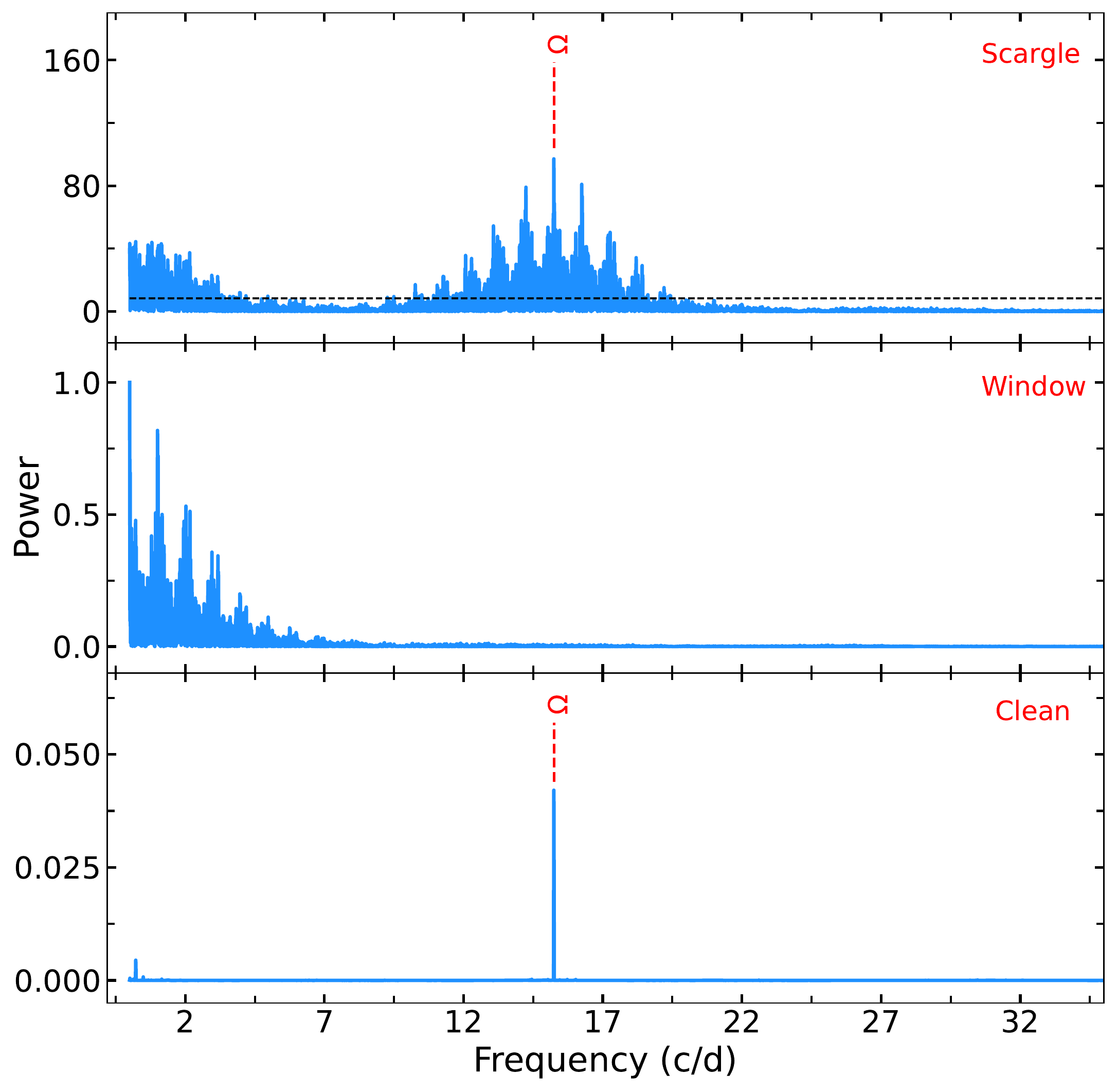} \label{fig:J1039_ps_opt}}
\subfigure[]{\includegraphics[width=0.45\textwidth]{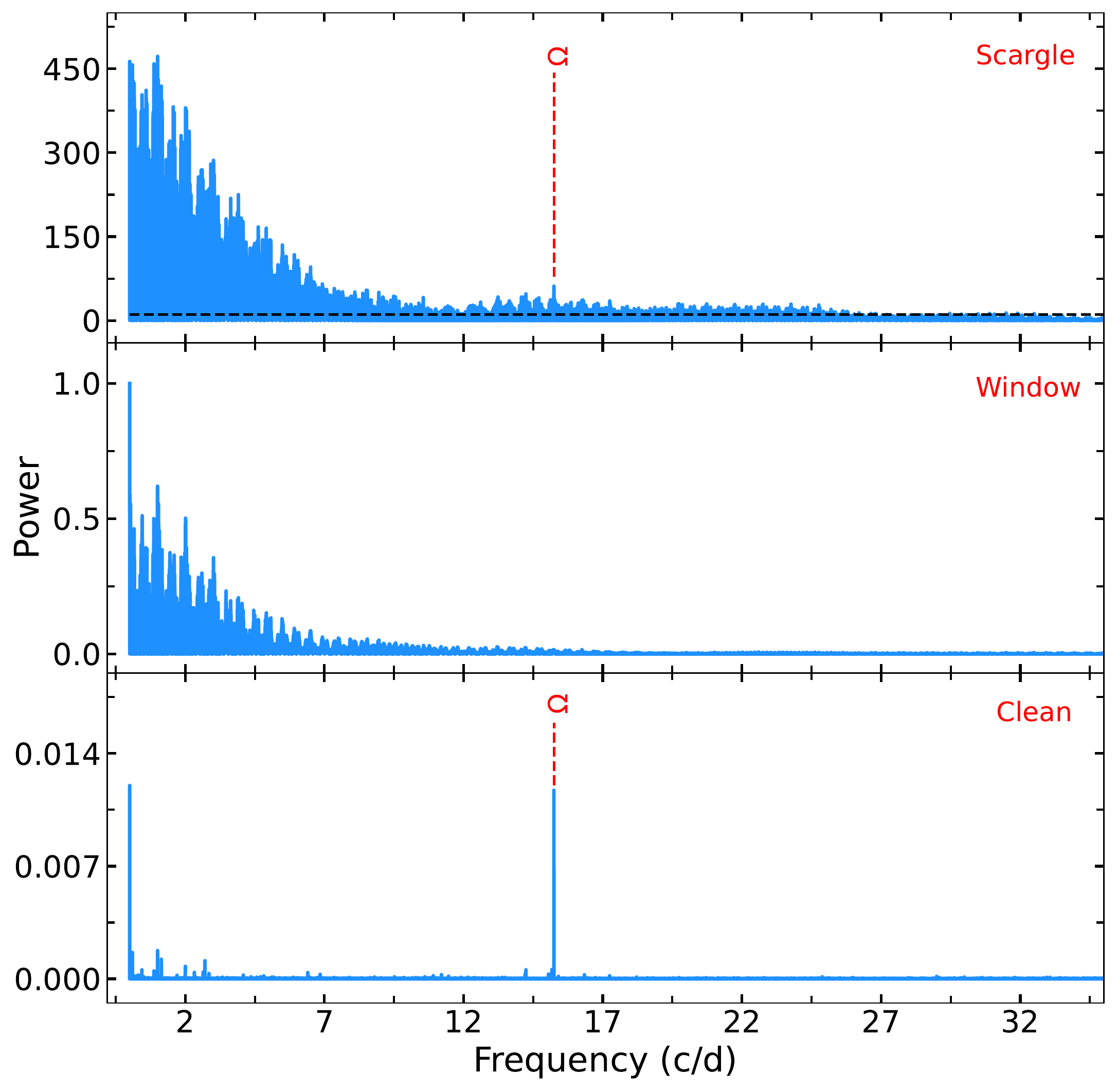} 
\label{fig:J1039_ps_aavso}}
\subfigure[]{\includegraphics[width=0.9\textwidth]{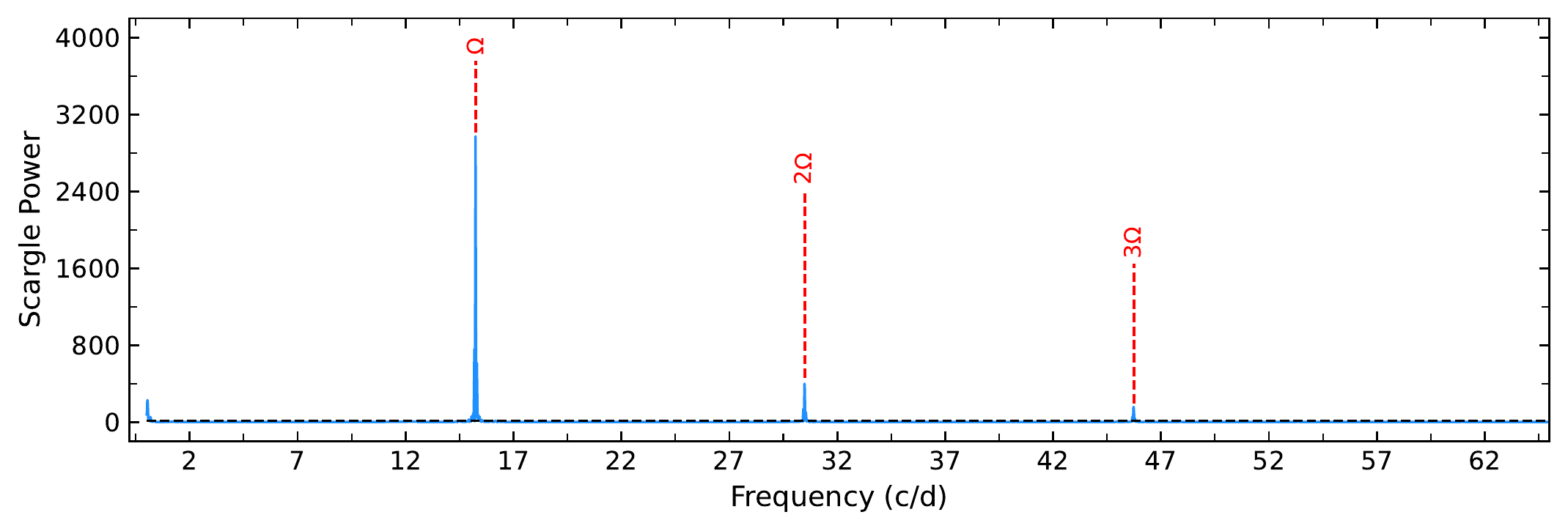} 
\label{fig:J1039_ps_tess}}
\caption{LSP (top panel), a window function (middle panel), and CLEANed power spectra (bottom panel) of YY Sex as obtained from (a) our optical, (b) AAVSO, and (c) \textit{TESS} observations. The horizontal dashed lines represent a 90 per cent significance level. }
\label{fig:J1039_ps}
\end{figure*}

% Figure-9
\begin{figure*}
\centering
\subfigure[]{\includegraphics[width=0.85\textwidth]{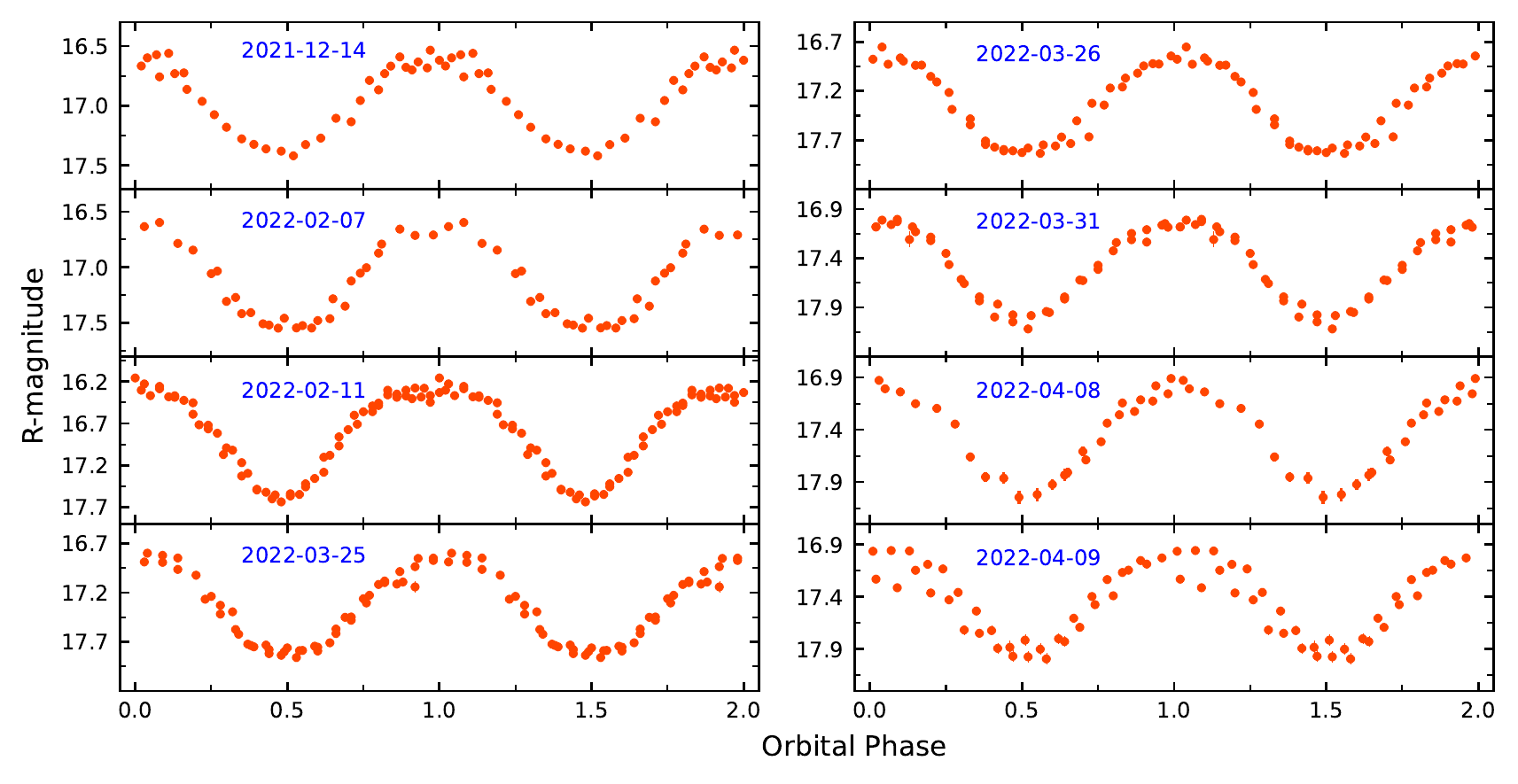} \label{fig:J1039_fold_opt}}
\subfigure[]{\includegraphics[width=0.85\textwidth]{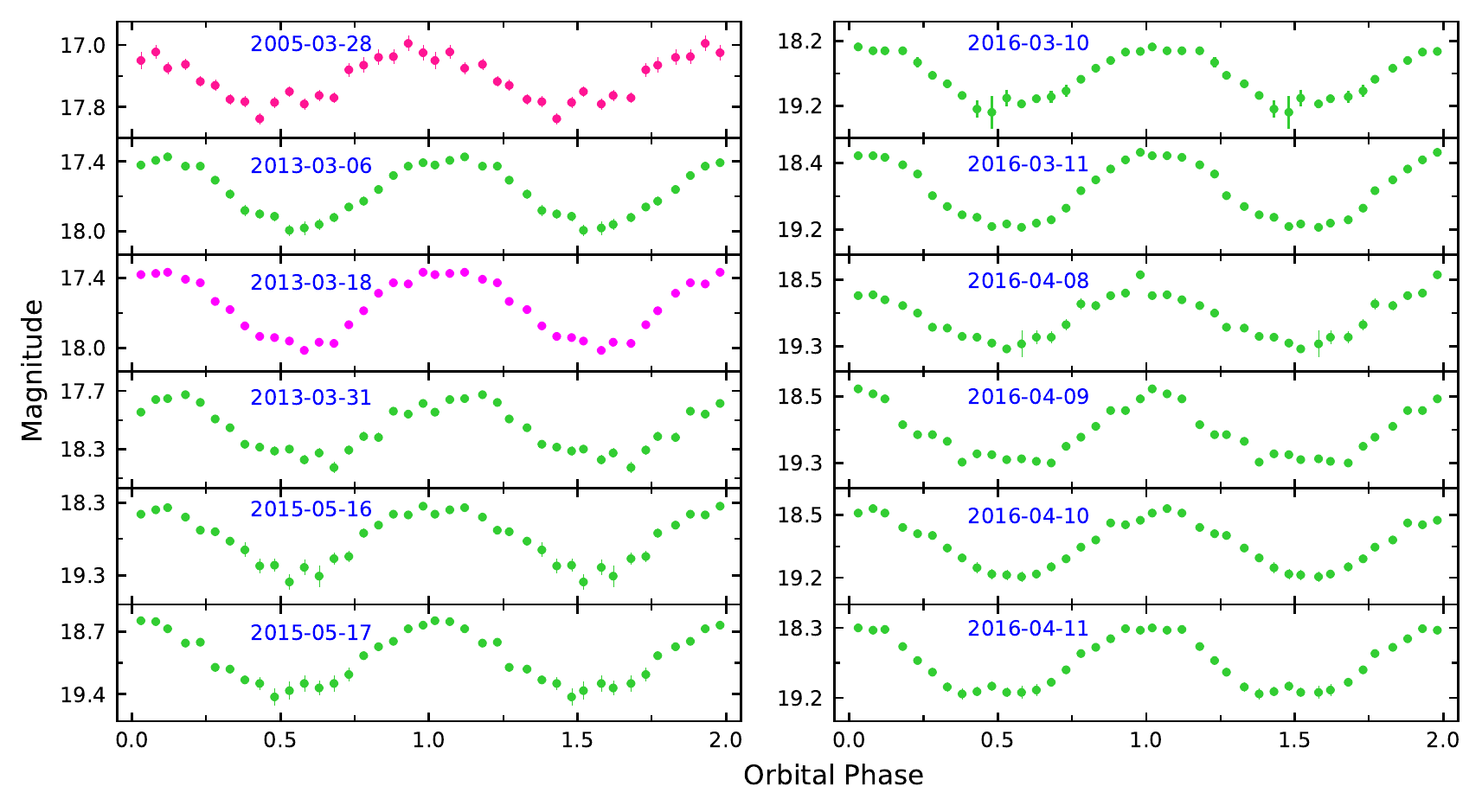}
\label{fig:J1039_fold_aavso}}
\subfigure[]{\includegraphics[width=0.82\textwidth]{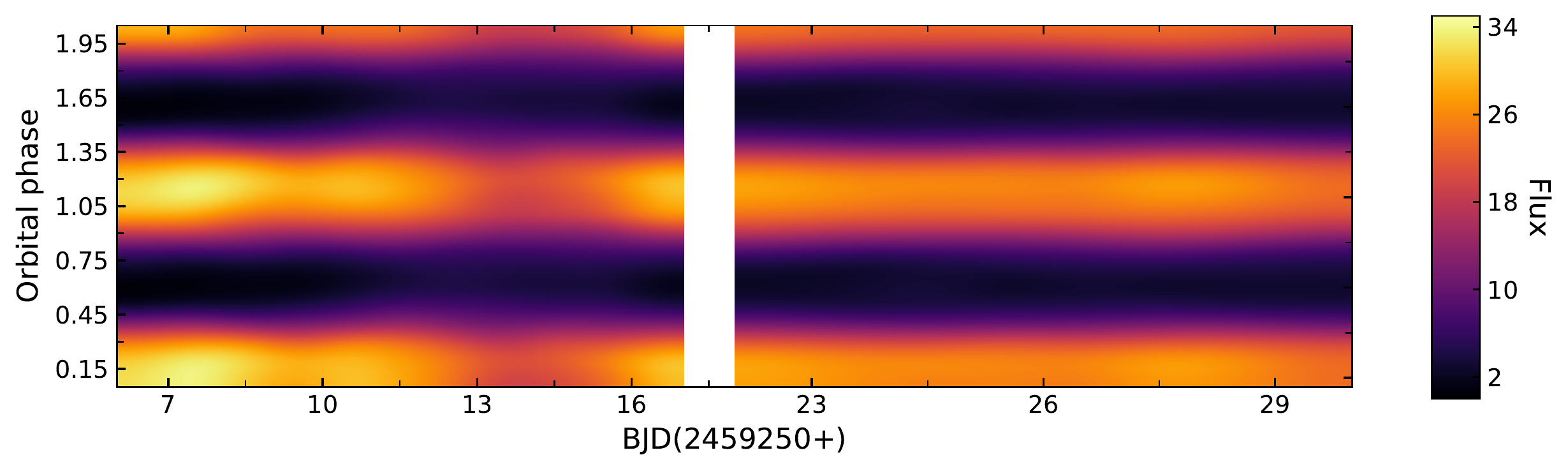} 
\label{fig:J1039_fold_tess}}
\caption{Folded light curve of YY Sex as obtained from (a) our optical R-band, (b) AAVSO, and (c) \textit{TESS} observations. For AAVSO observations, pink, green, and magenta colours indicate observations in CR, CV, and V bands.}
\label{fig:J1039_fold}
\end{figure*}

\subsection{YY Sex} \label{sec3.2}
\subsubsection{Optical photometry}
Fig. \ref{fig:J1039_lc_all} displays the long-term variable nature of the source using AAVSO-CR, AAVSO-CV, AAVSO-V, CRTS-V, and our R-band observations. The shaded region corresponds to the \textit{TESS} observations. Our optical-R band light curves of eight different epochs are shown in Fig. \ref{fig:J1039_lc_opt}, where the peak brightness of the system varies from 16.1 mag to 17.0 mag, indicating a change in the system's state. The light curves show sinusoidal-like variations with a variable peak-to-peak amplitude in the range of  $\sim$0.9-1.5 mag. The \textit{TESS} light curve is shown in Fig. \ref{fig:J1039_lc_tess}, where a clear and regular variation is observed.  

%Table-6
\begin{table}
\begin{center}
\caption{Maxima timings for YY Sex from AAVSO and our optical observations.}
\label{tab:maxima}
\end{center}
\begin{tabular}{cccc}
\hline
BJD & Cycle  & BJD & Cycle \\
\hline
2453458.3431(8) & 0  & 2457489.9861(5) & 61453 \\
2456358.8235(5) & 44211 & 2457490.0475(3) & 61454 \\
2456358.8872(7) & 44212 & 2459563.4351(4) & 93058 \\
2456370.1052(4) & 44383 & 2459618.2826(5) & 93894 \\
2456370.1735(5) & 44384 & 2459622.284(1) & 93955 \\
2456383.0978(4) & 44581 & 2459622.3492(4) & 93956 \\
2457160.9089(6) & 56437 & 2459622.4152(4) & 93957 \\
2457160.9756(4) & 56438 & 2459664.2727(3) & 94595 \\
2457458.0956(7) & 60967 & 2459664.3393(5) & 94596 \\
2457458.1671(4) & 60968 & 2459665.2567(4) & 94610 \\
2457459.1521(3) & 60983 & 2459665.3248(3) & 94611 \\
2457488.0158(5) & 61423 & 2459670.3093(3) & 94687 \\
2457488.0850(5) & 61424 & 2459678.1165(4) & 94806 \\
2457489.9216(4) & 61452 & 2459679.1682(4) & 94822 \\

\hline
\end{tabular}
\end{table}

% Figure-10
\begin{figure*}
\centering
\includegraphics[width=0.8\textwidth]{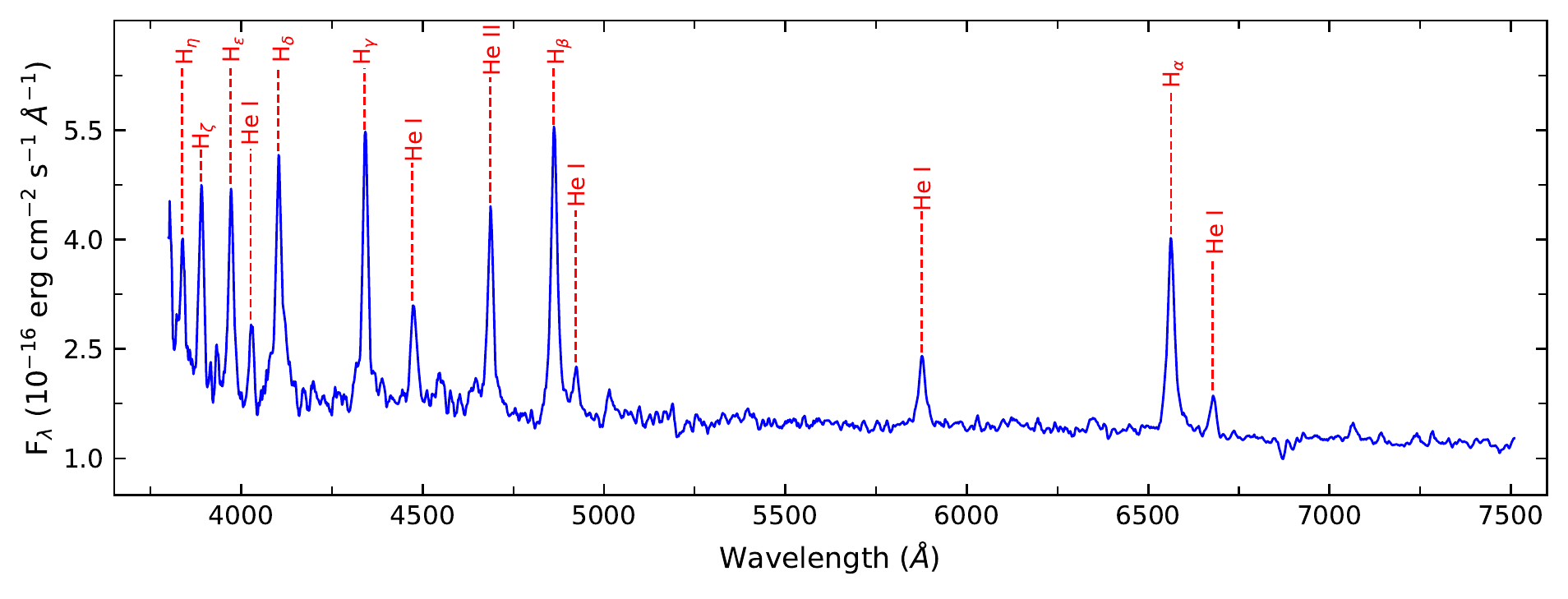}
\caption{Optical spectrum of YY Sex.}
\label{fig:J1039_spec_opt}
\end{figure*}

% Figure-11
\begin{figure*}
\centering
\includegraphics[width=0.9\textwidth]{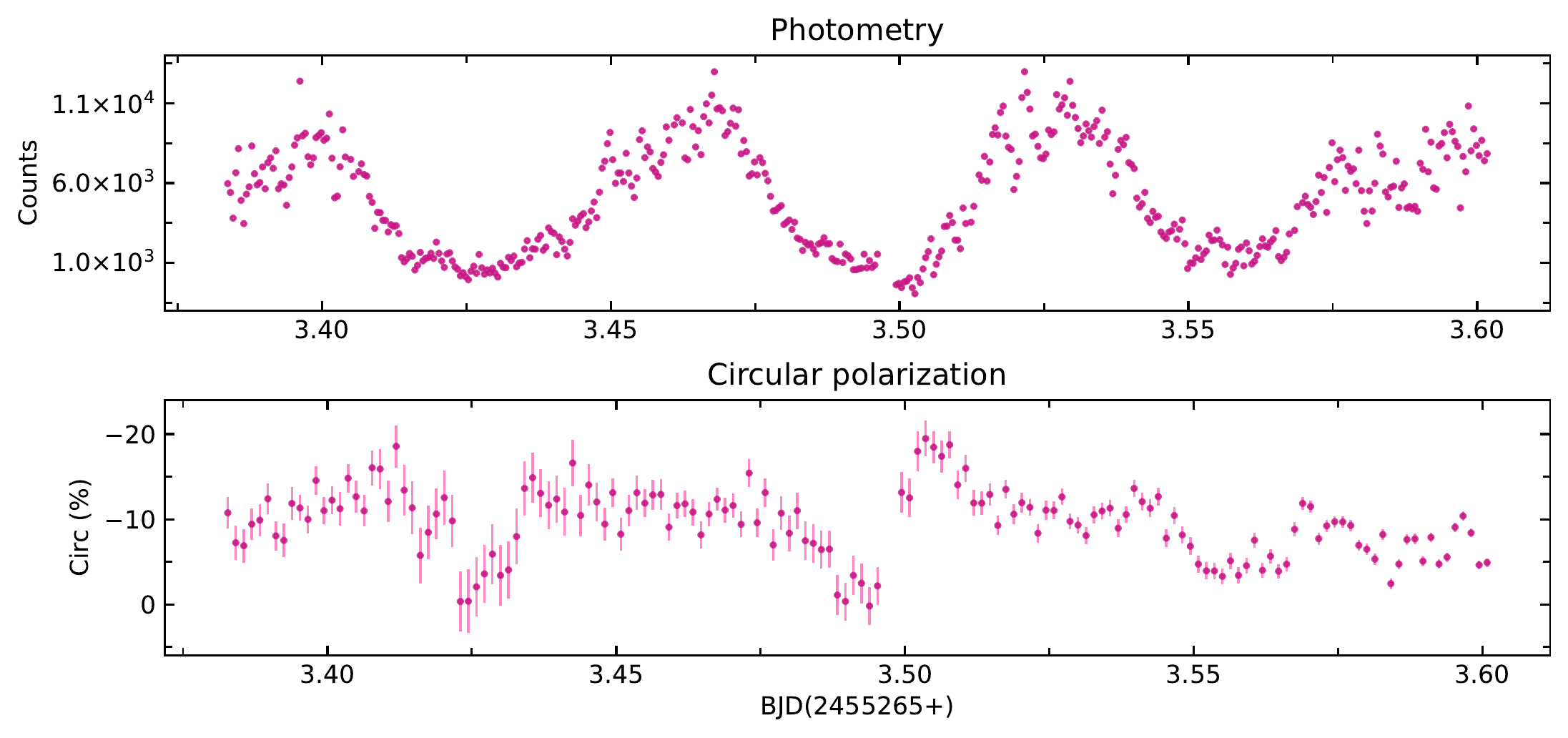}
\caption{Photometry and circular polarimetry of YY Sex. The upper and lower panel display photometric and circular polarised light curves, respectively.}
\label{fig:J1039_pol_opt}
\end{figure*}

%The observational feature of the reflection effect depicts a very large amplitude of the photometric modulation. \cite{2002PASP..114.1222W} also found the peak-to-peak magnitude range of 1.1 mag in their V-band light curves and reported that YY Sex has the largest amplitude modulations observed for CVs and detached white dwarf/M dwarf binaries \citep[see][]{1995MNRAS.275..100C}. 

% NOTE that the "reflection effect" and hot WD from the Warner and Woudt paper is now invalid given that we show that the large amplitude is of cyclotron origin and not a hot WD.

\par As discussed in Section \ref{sec3.1.1}, an LSP analysis was used to find the periodicity in the data. The LSP obtained from optical-R, AAVSO, and \textit{TESS} observations are shown in Fig. \ref{fig:J1039_ps}. The black horizontal dotted line represents the 90 per cent significance level. The LSP of our optical and AAVSO observations are noisy due to data gaps; therefore, as mentioned in Section \ref{sec3.1.1}, the CLEAN algorithm was employed with a loop gain of 0.1 and 1000 iterations. Only the \orb ~frequency was found to be present in the CLEANed spectra of R-band and AAVSO data, whereas we found \orb, \twoorb, and \threeorb ~frequencies in the LSP of \textit{TESS} data. The significant periods derived from optical-R, AAVSO, and \textit{TESS} data are given in Table \ref{tab:periods}. These values of the orbital period are well consistent with each other and also with the value reported by \cite{2002PASP..114.1222W}. We do not detect the presence of any additional frequencies, in particular, the proposed spin and beat frequencies of \cite{2002PASP..114.1222W}. 
\par We have determined 28 timings corresponding to maximum light using the optical-R and AAVSO data by fitting a Gaussian to the bright part of the light curve. The corresponding times are given in Table \ref{tab:maxima}. A linear fit between cycle numbers and maxima timings gave the following ephemeris for YY Sex:
\begin{equation} \label{eqn}
T_{0} = 2453458.346(1) + 0.06560523(2) \times E   
\end{equation}
where $T_{0}$ is defined as the time of maximum light and the errors are given in parenthesis.\\
%\textcolor{red}{is this a BJD ephemeris??? I don't understand table 5, how can the first be cycle 0 and the next cycle is 44211? also why are the cycles not integers?}
From Equation \ref{eqn}, we refined the orbital period to be 1.57452588 $\pm$ 0.00000002 h. Using the ephemeris provided in equation \ref{eqn}, the photometric light curves of all these observations were also folded and are displayed in Fig. \ref{fig:J1039_fold}. All of these folded light curves show broad maxima near phase 0.0 and minima at phase  $\sim$0.5 with variable brightness.

\subsubsection{Optical spectroscopy}
The optical spectrum of YY Sex is shown in Fig. \ref{fig:J1039_spec_opt}, where several emission features are visible. The spectrum exhibits strong single-peaked H-Balmer emission lines (H$\alpha$ to H$\eta$) and neutral and ionized He lines (He \rn{1} and He \rn{2}). Some of these lines were also seen in the earlier spectrum observed by \cite{2017ASPC..510..435G}. Based on the high intensity He \rn{2} 4686 \AA ~line compared to the hydrogen lines, \cite{2017ASPC..510..435G} reported that YY Sex belongs to the MCV category, possibly polars. Identification, fluxes, EW, and FWHM of lines observed in our optical spectrum are given in Table \ref{tab:opt_spec}. Balmer lines appear with an inverted Balmer-decrement of F(H$\alpha$)/F(H$\beta$)=0.71. We have also calculated the EW ratio of He \rn{2} 4686 \AA ~to {H$\beta$}, which comes out to be  $\sim$0.55. The strength of He \rn{2} 4686 \AA ~line is comparatively less than that of H$\beta$, as also seen in the observed spectrum of \cite{2017ASPC..510..435G}.

% Figure-12
\begin{figure*}
\centering
\subfigure[]{
\includegraphics[width=0.45\textwidth]{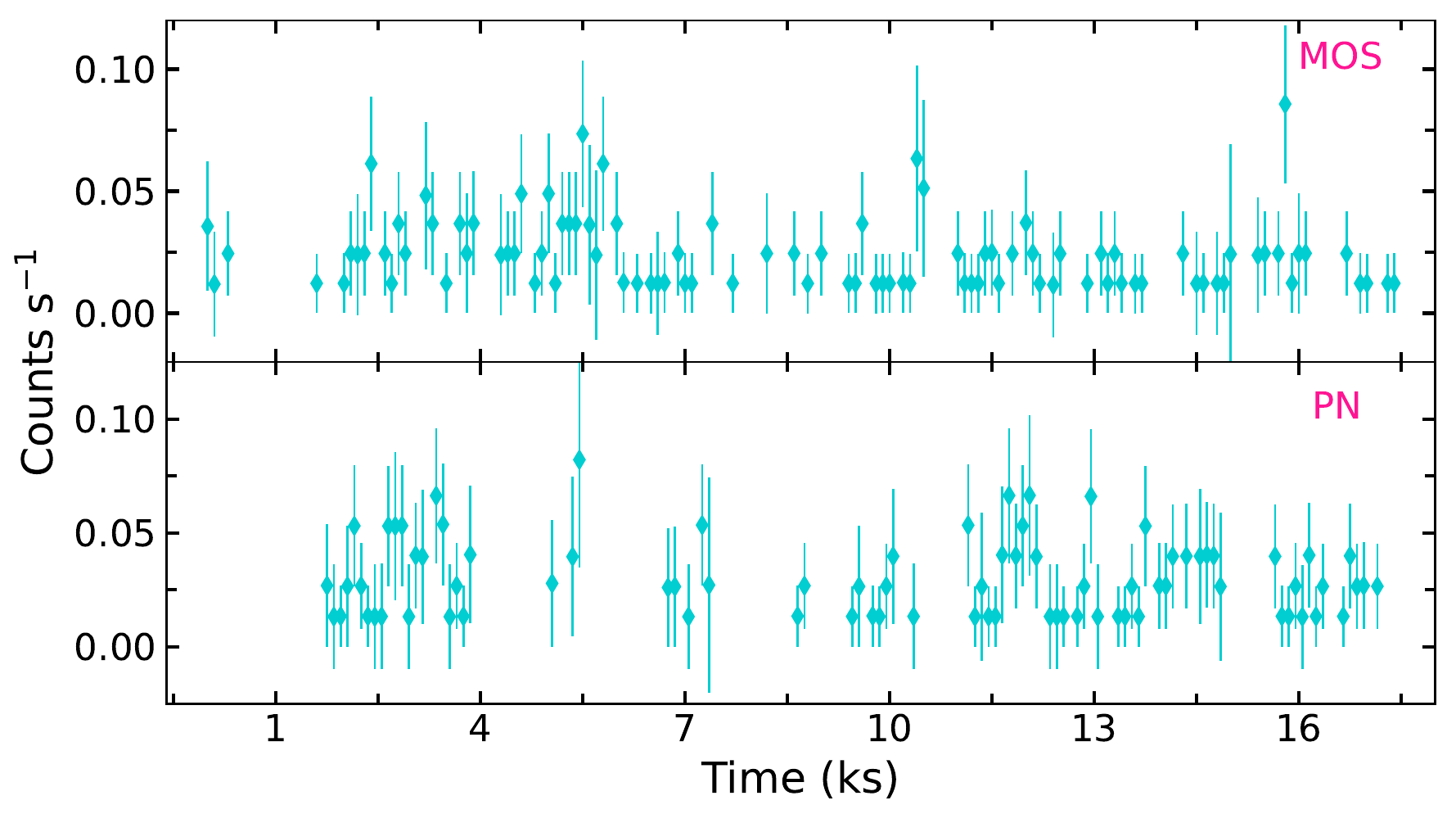}
\label{fig:J1039_lc_xray}}
\subfigure[]{
\includegraphics[width=0.45\textwidth]{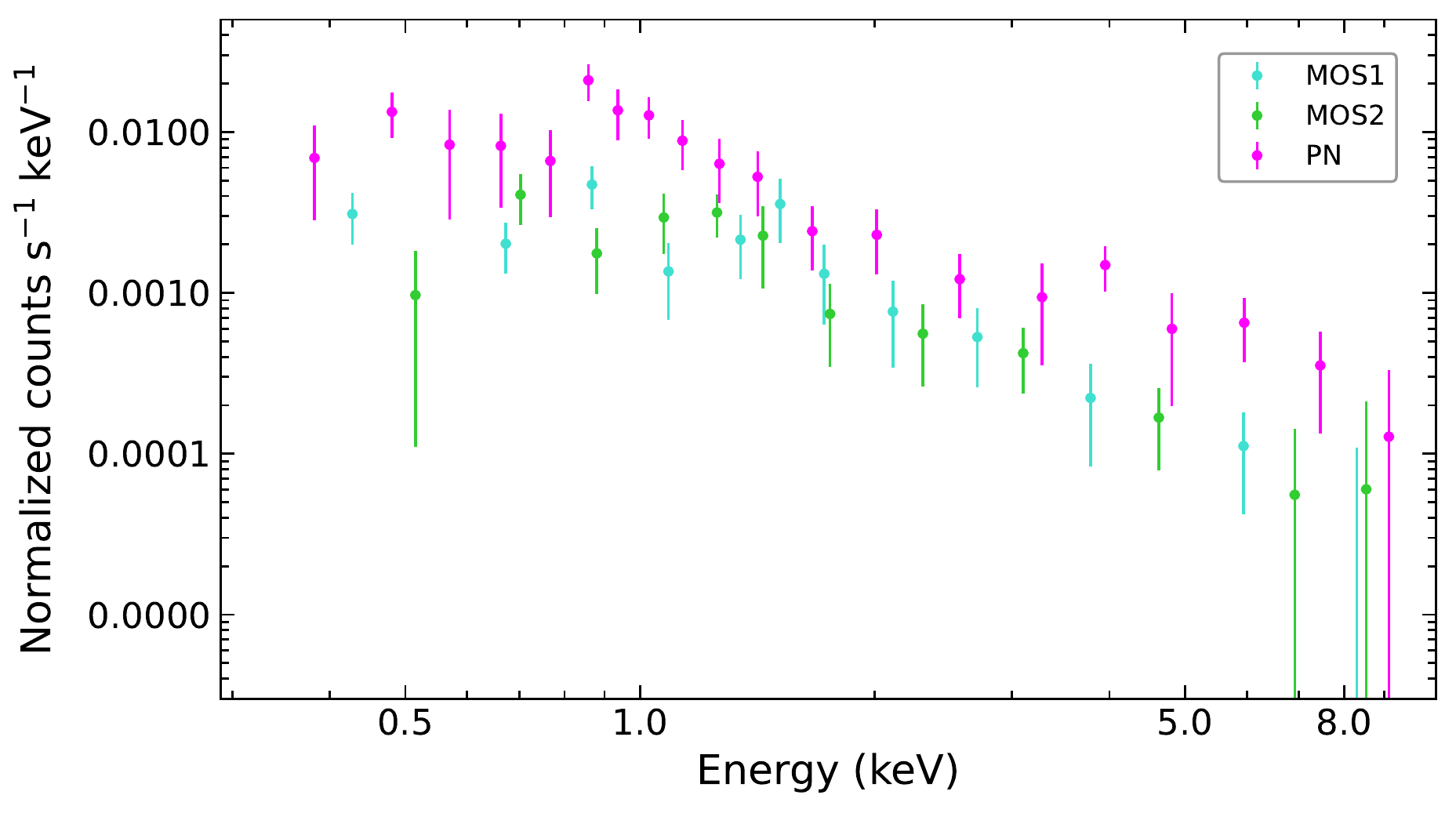}
\label{fig:J1039_spec_xray}}
\caption{(a) EPIC light curves of YY Sex in the 0.3-10.0 keV energy range binned in 50 s intervals. (b) X-ray spectra of YY Sex as obtained from MOS1, MOS2, and PN observations.}
\end{figure*}

\subsubsection{Polarimetry}
Fig. \ref{fig:J1039_pol_opt} displays the 5.76 h long, red filtered, photometric and circular polarimetric observations (upper and lower panels, respectively). The photometry displays the same orbital modulation as seen in our other optical data sets. In addition, the HIPPO photometry shows a relatively smooth light curve around the orbital minimum and a more variable light curve on the increase to the maximum. These additional features are not discernible in our other datasets due to their lower time resolution. This extra variability, on the rise to orbital photometric maximum, has been seen in other polars (including J1743 \citep{2012PZ.....32....3D} and e.g. \citep{Potter2010}) and in some cases, a period analysis reveals the variability to be quasi-periodic \citep{Potter2010}. We suggest that the frequencies detected in the analysis of \citet{2003MNRAS.339..731W} were probably of QPO in origin rather than spin and beat frequencies. Our polarimetric observations (see, Fig. \ref{fig:J1039_pol_opt}, lower panel) show the presence of circular polarization, modulated on the orbital period with an amplitude between 0 and -20 per cent. This unambiguously confirms that YY Sex is polar, showing the typical polarized variability as a result of the beaming of cyclotron emission from a single accretion region. The amplitude of the orbital photometric modulation can also be explained as a result of cyclotron emission rather than the proposed reflection effect of \citet{2003MNRAS.339..731W}.

\subsubsection{X-ray Timing and Spectra}
The background-subtracted X-ray light curves (with 50 s binning time) and spectra of YY Sex from EPIC-MOS and PN detectors in the 0.3-10.0 keV energy range are displayed in Figs \ref{fig:J1039_lc_xray} and \ref{fig:J1039_spec_xray}. Due to a poor signal-to-noise ratio, the power spectrum was found to be very noisy. Further, we also tried to fit the X-ray spectra with an appropriate model as described in Section \ref{sec3.1.2}. Unfortunately, the spectral parameters were also derived to be unphysical due to the poor signal-to-noise ratio. Therefore, we have not performed any further timing and spectral analyses.

%therefore we could not perform any further analysis. Further, the X-ray spectrum was also fitted with the appropriate model used for polars (see Section \ref{sec3.1.2}); however, due to the very poor signal-to-noise ratio,  the derived spectral parameters (\textbf{such as the mass accretion rate and bolometric flux}) were obtained to be unphysical (\textbf{an order of magnitude of two lesser than the default values}). 

\section{Discussion} \label{sec4}
We have carried out detailed timing and spectral analyses of two candidate MCVs, J1743 and YY Sex, using optical and X-ray data. For both sources, we confirm and refine the previously determined orbital periods. The orbital periods of J1743 and YY Sex populate them at the lower edge and below the period gap of CVs, respectively. Based on the present analyses, we confirm that both objects belong to the polar subclass of MCVs. The key observational features which secure the identification of these systems as polars are given in Table \ref{tab:comparison_table}.

%\textbf{We confirm that both objects belong to the polar subclass of MCVs based on the majority of observational features (see  Table \ref{tab:comparison_table}, for details).}

% Table-7
\begin{table}
\renewcommand{\arraystretch}{1.2}
	\centering
\caption {Key observational characteristics of polars obtained for J1743 and YY Sex.} \label{tab:comparison_table}
\begin{tabular}{p{0.5\columnwidth}p{0.2\columnwidth}p{0.2\columnwidth}}
\hline
Observational features & J1743 & YY Sex \\
\hline
Only one periodicity in optical &  Yes & Yes \\
Only one periodicity in X-rays &  Yes & NA \\
EW[He \rn{2} 4686 \AA]/EW[{H$\beta$}] > 0.4 & Yes & Yes \\
EW[{H$\beta$}] > 20 \AA & Yes & Yes \\
Inverse Balmer decrement & Yes & Yes \\
Strong circular polarization & NA & Yes \\
Soft blackbody emission & Yes & NA \\
Luminosity (<10$^{33}$ erg s$^{-1}$) & Yes & NA \\

\hline
\end{tabular}

\bigskip
\emph{\textbf{Note.} NA indicates that the data is not available to comment on anything.} 
\end{table}

\par From multiple epoch optical observations of J1743, we have found indications that its accretion rate varies on timescales of days, which suggests that it could be due to the variable strength of a constant stream. However, for those epochs where rapid variability is seen, it could be due to `blobs' of material being transferred. J1743 was observed in a high state between March and May 2021 before dimming significantly in June and brightening again by April 2022. Such high and low states are one of the typical characteristics of polars \citep{1990SSRv...54..195C}.    The soft and complex double-hump X-ray light curve is seen for J1743, where a broad prominent hump is generally associated with the main accretion region. The second hump, on the other hand, indicates the presence of an independent second accretion region or accretion at a second pole.
%However, the second hump indicates the presence of an independent second accretion region or accretion at a second pole. 
The UV light curve also displays a double-hump profile but with a shift with respect to the X-ray light curves.  There have been few polars where accretion onto a second pole has been indicated, e.g., DP Leo \citep{1990SSRv...54..195C}, VV Pup \citep{1989ApJ...342L..35W}, UZ For \citep{1990A&A...230..120S}, QS Tel \citep{1995A&A...293..764S}, MT Dra \citep{2002A&A...392..505S}, AM Her \citep{1985A&A...148L..14H, 2020A&A...642A.134S}, and V496 Uma \citep{2022A&A...662A.116O, 2022MNRAS.513.2930K}, etc.  Therefore, there is a need for polarimetric observations of J1743, which can help to confirm the two-pole accretion. The complex soft X-ray light curves have also been observed in polars QQ Vul, EF Eri, AN Uma, and V834 Cen \citep{1985SSRv...40...99M}. Furthermore, the light curves of two sources, QQ Vul and V834 Cen, seem to have drastically changed in later observations \citep{1987ApJ...315L.123O, 1994ApJ...424..947S}. Even if the WD has a simple, centred, dipole field, the interaction of the magnetic field and the ballistic stream can lead to an elongated accretion spot or spots \citep{1988MNRAS.232..175M, 2002PhDT.........7C}. Therefore, apart from the second accretion region or two-pole accretion, a double-spot structure with an elongated main spot could generate such complex double-humped soft X-ray light curves \citep{2017PASP..129f2001M}. The mass transfer rate can further change the spot geometry and the light curve morphology, as we have seen in our optical light curves. Further, the light curve variations of YY Sex show the presence of only one frequency (the orbital frequency) and its harmonics in the power spectra, confirming that it is a polar rather than an IP, as suggested by \cite{2002PASP..114.1222W}. In addition, our observations show circular polarization modulated on the orbital period, thereby unambiguously identifying YY Sex as a polar. This also explains the large amplitude photometric variation as being the result of the changing viewing angle, to the beamed cyclotron emission, over the orbital period, which also rules out the scenario of a large reflection effect from a hot WD as indicated by \cite{2002PASP..114.1222W}. Using the mean empirical mass-period relation of \cite{1998MNRAS.301..767S}, we have estimated the mass and radius of the secondary of J1743 and YY Sex as 0.15 $\pm$ 0.02 M$\odot$, 0.20 $\pm$ 0.01 R$\odot$ and 0.09 $\pm$ 0.02 M$\odot$, 0.14 $\pm$ 0.01 R$\odot$, respectively. We have also estimated the mean density of the secondary in J1743 and YY Sex as 24.8 g cm$^{-3}$ and 43.2 g cm$^{-3}$ \citep[see][for formulae]{1995cvs..book.....W}, respectively. For the derived  orbital periods of J1743 and YY Sex, secondaries correspond to lower main-sequence stars of spectral classes M4.2 and M6.2, and effective temperatures of 3260 K and 2823 K, respectively \citep{2006MNRAS.373..484K}.

%These light curves in terms of either amplitude or shape resemble with some well-studied polars, e.g., QQ Vul \citep{1984ApJ...277..682N}, V834 Cen \citep{1986MNRAS.218..201C}, V1500 Cyg \citep{1988ApJ...332..287K}, etc. \cite{2002PASP..114.1222W} also suspected that these features are characteristics of the recent nova in which either the accretion disc does not dominate, or it is completely absent, as seen in polars on the basis of classical nova model given by \cite{1986ApJ...310..222P}.  Furthermore, V1500 Cyg has also shown a modulation range of 1.2 mag in the year 1987 (12 years after the outburst), which reduced to 0.7 mag in the year 1995 \citep{1999A&A...352..563S}.  However, in the case of YY Sex, the modulation has not reduced even after 20 years since 2002 after the first reported value of modulation by \cite{2002PASP..114.1222W}.  THIS DISCUSSION IS obsolete and not relevant given our detection of polarization.

\par The optical spectra of J1743 and YY Sex are very similar to the spectra of other polars. The most evident spectral signature of mass accretion is the presence of Hydrogen Balmer, He \rn{1}, and He \rn {2} emission lines. \cite{1992PhDT.......119S} provided a criterion to classify MCVs on the basis of the emission lines in the optical spectra. According to Silber's criteria, MCVs are characterized by an EW ratio of He \rn{2} 4686 \AA/H$\beta$ > 0.4 and a large EW of H$\beta$ (> 20 \AA). We have detected a large value of EW of H$\beta$ for both sources. Also, the EW ratio of He \rn{2} 4686 \AA/H$\beta$ comes out to be > 0.4, suggesting that these two sources have a magnetic nature. Further, the Balmer decrement H$\alpha$/H$\beta$ was estimated to be 0.86 and 0.71 for J1743 and YY Sex, respectively. The ratio is inverted compared to the recombination spectrum from an optically thin gas \citep{1971MNRAS.153..471B}. Optical thickness alone can not account for the ratio of H$\alpha$/H$\beta$. Therefore, collisions between levels n=2 and n=3 are important, which gives a lower limit to the electron density N$_{e}$ > 10$^{12}$ cm$^{-3}$ \citep{1977ApJ...217..815S, 1990SSRv...54..195C}. All these features indicate that the lower Balmer lines may arise in collision-dominated plasma in an optically thick medium. 

\par The X-ray spectra of J1743 reveal a multi-temperature post-shock region with T$_{\rm low}$ < 840 eV and T$_{\rm high}$  $\sim$32 keV. However, the multi-temperature plasma produced in the post-shock region requires the presence of two absorbers; a thin absorber of 1.79 $\times$ 10$^{21}$ cm$^{-2}$ and a thick absorber of  $\sim$7.5 $\times$ 10$^{23}$ cm$^{-2}$ covering  $\sim$56 per cent of the X-ray source. A blackbody temperature of  $\sim$97 eV  was also found to be present, indicating that a fraction of the hard X-rays is reprocessed and re-radiated in the soft X-rays. However, the blackbody temperature was found to be relatively high, unlike other polars AM Her \citep{2020A&A...642A.134S}, AI Tri \citep{2010A&A...516A..76T}, QS Tel \citep{2011A&A...529A.116T}, V496 Uma \citep{2022A&A...662A.116O}, etc. %Such a high value of kT$_{\rm bb}$ was also found in eclipsing polar VV Pup \citep{2020A&A...633A.145B}. 
The softness ratio, F$_{\rm s}$ /4F$_{\rm h}$ \citep[see][for details]{2004MNRAS.347..497R} was then calculated and found to be  $\sim$0.01, which rejects the presence of any soft X-ray excess. We have presumed that both the soft and hard components were subjected to the same partial covering absorbers, while estimating the value of F$_{s}$ /4F$_{h}$. \cite{2004MNRAS.347..497R} found large soft X-ray excess only in 13 per cent of  their sample. They also found numerous polars without a discernible soft component implying that the presence of a soft X-ray excess could not be associated with the primary characteristic of polars. With a mass accretion rate of  $\sim$5 $\times$ 10$^{-12}$ M$\odot$ yr$^{-1}$ and bolometric luminosity of  $\sim$4 $\times$ 10$^{31}$ erg s$^{-1}$, the parameters of J1743 fall in the typical range of polars.  Assuming T$_{\rm high}$  $\sim$32 keV as the maximum post-shock temperature and using the mass-radius relationship of \cite{1972ApJ...175..417N}, we have derived the WD mass to be 0.75$_{-0.11}^{+0.16}$ M$\odot$ for J1743.

\section{Conclusions} \label{sec5}
 Based on the optical and X-ray timing and spectral analyses, we  conclude that these MCV candidates, J1743 and YY Sex belong to the polar subclass of MCVs. We summarize our findings as follows:
\begin{itemize}
    \item[1.] We confirm and refine the orbital period of J1743 using the present analysis. It has shown brightness variation between low and high states. The presence of strong He \rn{2} 4686 \AA ~and H$\beta$ lines in the optical spectrum indicates the magnetic nature of the system. The multi-temperature post-shock region is absorbed through thin and thick absorbers. A fraction of the hard X-rays is reprocessed and re-radiated in the soft X-rays; however, we did not find any evidence of soft X-ray excess in this source. We have derived a mass accretion rate of  $\sim$5 $\times$ 10$^{-12}$ M$\odot$ yr$^{-1}$, bolometric luminosity of  $\sim$4 $\times$ 10$^{31}$ erg s$^{-1}$, WD mass of 0.75$_{-0.11}^{+0.16}$ M$\odot$ for J1743. All the above-mentioned features with the present data confirm that J1743 is indeed a polar.

    \item[2.] Similar to J1743, we confirm and refine the orbital period of YY Sex using the present analysis. The presence of several emission lines and strong hydrogen Balmer lines with strong He \rn{2} 4686 \AA ~and H$\beta$ confirms the magnetic nature of accretion flow. The low and high states in the long-term light curve, the presence of only one period and its harmonics, the detection of orbitally modulated circular polarization, and the emission line features in the optical spectrum confirm that YY Sex belongs to the category of polars.
\end{itemize}

\section{Acknowledgements}
We thank the anonymous referee for providing helpful suggestions. The observing staff and observing assistants of 1-m class telescopes are deeply acknowledged for their support during optical photometric observations. We thank the staff of IAO, Hanle and CREST, Hosakote, that made spectroscopic observations possible. The facilities at IAO and CREST are operated by the Indian Institute of Astrophysics, Bangalore. NR acknowledges Mr Vibhore Negi for his help in various technical aspects. ASH and JCP thank the Ministry of Innovation Development of Uzbekistan and the Ministry of Science and Technology of India for financing the joint project (Project References: UZB-Ind-2021-99 and INT/UZBEK/P-19). This research has made use of the data obtained with \textit{XMM-Newton}, an ESA science mission with instruments and contributions directly funded by ESA Member States and NASA. This paper also includes data collected with the \textit{TESS} mission, obtained from the MAST data archive at the Space Telescope Science Institute (STScI). Funding for the \textit{TESS} mission is provided by the NASA Explorer Program. We acknowledge with thanks the variable star observations from the AAVSO International Database contributed by observers worldwide and used in this research. The CRTS survey is supported by the U.S.~National Science Foundation under grants AST-0909182 and AST-1313422. 

\section{DATA AVAILABILITY}
The \textit{XMM-Newton} data used for analysis in this article are publicly available in NASA’s High Energy Astrophysics Science Archive Research Center (HEASARC) archive (\url{https://heasarc.gsfc.nasa.gov/docs/archive.html}). The \textit{TESS} data sets are publicly available in the \textit{TESS} data archive at \url{https://archive.stsci.edu/missions-and-data/tess}. The AAVSO, ASAS-SN, and CRTS data sets are available at \url{https://www.aavso.org/data-download} and  \url{https://asas-sn.osu.edu/variables}, and \url{http://nunuku.caltech.edu/cgi-bin/getcssconedbid_release2.cgi}, respectively. The optical photometric and spectroscopic data underlying
this article will be shared on reasonable request to the corresponding
author.

\bibliographystyle{mnras}
\bibliography{optical.bib}

\appendix

\label{lastpage}
\end{document}